\documentstyle[preprint,tighten,aps]{revtex} 
\begin{document}           %
\draft
\preprint{\vbox{\noindent 
To appear in Nuclear Physics B\hfill hep-ph/9511383\\
          \null\hfill MIT-CTP\#2361\\ 
          \null\hfill INFNCA-TH-94-25}}
\title{The Dimensionally Reduced Effective Theory for Quarks\\
 in High Temperature QCD}
\author{Suzhou Huang\cite{email1}}
\address{
Center for Theoretical Physics, Laboratory for Nuclear Science
and Department of Physics, \\
Massachusetts Institute of Technology, Cambridge, 
Massachusetts 02139;\cite{present}\\
and Department of Physics, FM-15, University of Washington,
Seattle, Washington 98195
        }
\author{Marcello Lissia\cite{email2}}
\address{
Center for Theoretical Physics, Laboratory for Nuclear Science
and Department of Physics, \\
Massachusetts Institute of Technology, Cambridge,
Massachusetts 02139;\\
and Istituto Nazionale di Fisica Nucleare, 
via Ada Negri 18, I-09127 Cagliari, Italy;\cite{present}\\
and Dipartimento di Fisica dell'Universit\`a di Cagliari, I-09124 Cagliari,
Italy
        }
%
\date{November 1995; revised May 1996}
\maketitle                 
\begin{abstract}
  We show that QCD undergoes dimensional reduction 
at high temperatures also in the quark sector. In the kinematic
region relevant to screening physics, where the lowest Matsubara
modes are close to their ``mass-shells'', all static  Green's
functions involving both quarks and gluons, are reproducible in
the high-$T$ limit by a renormalizable three dimensional Lagrangian
up to order $\tilde{g}^2(T)\sim 1/\ln T$. This three dimensional theory 
only contains explicitly the lightest bosonic and fermionic Matsubara 
modes, while the heavier modes correct the tree-level couplings and 
generate extra local vertices. We also find that the quark degrees of 
freedom that have been retained in the reduced theory are
nonrelativistic in the high-$T$ limit. We then improve our result
to order $\tilde{g}^4(T)$ through an explicit nonrelativistic
expansion, in the spirit of the heavy quark effective theory.
This effective theory is relevant for studying QCD screening 
phenomena with observables made from quarks, e.g. mesonic and baryonic 
currents, already at temperatures not much higher than the chiral 
transition temperature $T_c$.
\end{abstract}
\narrowtext
\maketitle
\section{Introduction}
\label{intro}
   The behavior of field theories, quantum chromodynamics (QCD) in
particular, at finite temperature ($T$) 
is of great phenomenological and theoretical interest. 
In general, we can roughly classify finite temperature physics into
two categories: real-time dynamics and screening phenomena. On one hand,
real-time dynamics describes the time-dependent response of a system
(correlators as functions of real frequency) to time-dependent external 
probes. On the other hand, screening phenomena refer to the static 
spatial-dependent response (correlators as functions of the spatial 
momentum) to time-independent external probes.
 
   Without doubts, real-time dynamics is highly interesting and important,
and it could apply, in principle, to a wide range of phenomenological 
applications. Unfortunately, its complexity and the present lack of
systematic nonperturbative approaches make real-time dynamics at finite 
temperature a territory of field theory that is still, to a very  large
extent, barely cultivated~\cite{realt}. In addition, the difficulties
of realizing an equilibrated experiment with a given temperature,
at least in the specific case of QCD, has practically prevented us from
accessing real-time data that are not tempered by some {\em ad hoc}
phenomenological assumptions.

   In contrast, the Euclidean nature of the static correlation functions
makes the physics associated with screening phenomena relatively
simpler. In fact, static correlation functions are determined from 
equilibrium ensembles and involve no tricky analytic 
continuation~\cite{realt,contra} and, therefore, screening physics is
well-suited to the lattice formulation of field theories at finite
temperature. As a consequence, lattice QCD at finite temperature provides
us with a large body of measurements not only of bulk quantities, such 
as the specific heat and pressure, but also of more detailed observables
ranging from screening masses~\cite{mass-s} to screening
wave functions~\cite{bs-s}.

   In the past few years several physical pictures or scenarios have been
proposed for properly understanding and interpreting the available 
lattice data, both the data involving observables made from pure gluonic
fields~\cite{reisz} and the data involving observables made from explicit 
quark fields~\cite{bs-lattice}. One of the most important concepts used
in these works is the so-called dimensional reduction (DR) at high
temperatures: this concept can be roughly summarized by saying that
these QCD screening observables can be effectively described by a
three-dimensional theory when temperature is high enough.

   The DR picture is based on the observation~\cite{ginsparg,Appelquist81} 
that two different scales appear, in general, in field theories at high~$T$:
one scale is order $T$ and the other is order one relative to $T$ (this
second scale is strictly order one, i.e. independent of $T$, only at the 
tree-level in general).
At high $T$ these two scales become vastly separated and certain degrees 
of freedom effectively decouple; this phenomenon is analogous to the 
decoupling of heavy particles~\cite{Appelquist75}. So far the existing
literature~\cite{Nadkarni83,landsman,kajantie} has mostly concentrated on
situations where the observables involved are purely bosonic: the zero
Matsubara modes (scale of order one) are the explicit light degrees of
freedom, while the non-zero modes (scale of order $T$)
play the role of the heavy degrees of freedom.

    However, we are often interested in observables that couple directly
to quark degrees of freedom, observables that would vanish without the
explicit presence of quarks in the theory. Our interest in this kind of
observables is by no means academic, but rather it is dictated by the
fact that many important observables fall into this class, such as
mesonic and baryonic correlators, and in fact such observables have been
extensively studied at $T=0$~\cite{shuryak}. However, less attention has
been paid up to now to such correlators between observables made from quark
fields in the context of DR. The reason is probably that the scale
separation in these situations is less clear, since all fermionic
Matsubara modes have energies of order $T$, due to their antiperiodic
boundary condition. Therefore, only the specific underlying dynamics can
make DR possible for observables that are made directly with quarks. In
other words, it is the theory itself that must generate the scale
separation, a necessary condition for the decoupling of some degrees
of freedom.

    In a recent letter~\cite{drl} we were able to show in a specific
asymptotically free theory, the Gross-Neveu model, that a new scale
smaller than $T$, i.e. $T/\ln T$, naturally emerges at high $T$. It
is this dynamically generated scale that provides the scale hierarchy 
that in turn makes DR possible for fermionic observables in that model. 
The purpose of this paper is to examine what happens to QCD at high
temperature, to verify that QCD undergoes DR to one-loop order also in 
the quark sector, and to give the corresponding explicit form of the 
reduced theory.

    As we will show, the DR in the quark sector is not
based on the fact that the lightest Matsubara mode $(\pi T)$ is much
lighter than the next lightest mode $(3\pi T)$. Instead, the small
expansion parameter is the so-called off-shellness or residual momentum
after subtracting $\pi T$. This expansion parameter roughly measures how
far is a given configuration from the noninteracting valence state and
its meaning can be clarified by considering the example
of the Hydrogen atom. One can usually ignore the two-electron-one-positron
configuration relative to the one with a single electron, not because 
two electrons plus one positron are much heavier than one single electron, 
but rather due to the fact that the two-electron-one-positron configuration
is about $1.5$ MeV ($3 m_e$) off the energy-shell of the free state, 
whereas the off-shellness of single-electron configuration is only due to
the binding energy, $\sim 13$~eV.

    In the most strict sense DR requires that the reduced theory be
renormalizable in three-dimensions and that corrections to correlations
calculated in this reduced theory be suppressed by powers of $1/T$. 

    It is well-known that only the first few terms of the reduced
theory are renormalizable in three-dimensions, even when all degrees
of freedom are bosonic~\cite{landsman}. In our particular case,
we also find a three-dimensional renormalizable Lagrangian only up 
to one-loop order.
Nonetheless, we can still derive an effective theory that is
capable of describing screening physics with an accuracy better
than the leading order. This kind of reduced theory in general
contains higher dimensional operators and needs to be defined in
some well-defined regularization scheme. The coefficients of the
reduced Lagrangian are determined by requiring that, in the appropriate
kinematic regime, the relevant one-particle irreducible graphs calculated 
in the reduced theory match the corresponding ones in the original theory.
It is important to understand that, even though the coefficients 
are calculated perturbatively, the reduced theory is designed to maintain
the infrared properties of the original theory and, hence, the solution
of the reduced theory is in general nonperturbative~\cite{Gross81}.

    In the Gross-Neveu model~\cite{drl} we have explicitly shown that the
fermionic degrees of freedom that survive in the reduced Lagrangian are
nonrelativistic. Futhermore, it has been suggested~\cite{bs-lattice} 
that the same happens to the quark degrees of freedom present in the
QCD Lagrangian at high $T$. In this paper, we demonstrate that this is
in fact the case for QCD and hence derive the nonrelativistic Lagrangian
for the quarks up to one loop with methods similar to those used in
deriving the heavy-quark effective theory~\cite{caswell,lepage}. This
effective Lagrangian should reproduce screening mass splittings up to
order $\tilde{g}^4(T)$ and masses themselves up to order $\tilde{g}^2(T)$.

    We want to point out from the outset that our effective Lagrangian
is relevant for describing the long-distance behavior of spatial 
correlators only in those channels that are protected by some global 
symmetry (such as baryon number and flavor) from mixing with gluons,
e.g., baryonic and mesonic screening correlators.
Furthermore, this reduced theory is not meant for describing 
correlators that can mix with gluonic state (their large distance 
behavior is dominated by screening glueball masses), thermodynamical 
potentials or other bulk quantities that are dominated by the bosonic 
zero modes. In these later cases, the terms in the effective 
Lagrangian containing fermions give irrelevant contributions and can 
be dropped.

    The plan of the paper is the following. 

In the next section we outline the general strategy and the criteria 
for dimensional reduction when observables made with quarks are involved. 
Since the concepts we introduce in Sec.~\ref{DR} are somewhat new and 
cannot be explicitly found in the existing literature, we make the discussion
as complete as possible. In particular, we first review the screening
physics and the observables the effective theory is intended do describe,
e.g. spatial correlators, screening masses and other observables 
that are identically zero when quarks are explicitly ignored. The 
relevant kinematic region and the exact meaning of the mass-shell 
condition in the context of screening physics are specified next. 
We finally introduce the expansion parameter, the off-shellness, in 
the same section.

In Sec.~\ref{DRcalc} we explicitly calculate the DR Lagrangian, both at 
the tree and one-loop levels in the specified kinematic region.
Composite operators are considered in Sec.~\ref{CO}.

In Sec.~\ref{NRred} we first discuss a power counting scheme that
guarantees that the expansion parameter remains small and, therefore,
that the expansion is self-consistent.
Then we derive an effective Lagrangian for QCD, where
quarks are treated nonrelativistically, in analogy with heavy
quarkonium systems. The final form of the reduced Lagrangian and the
relationships between parameters in QCD and in the reduced theory are
explicitly given. We finally discuss the temperature regime
where the reduced theory becomes quantitatively reliable.

The final section contains the summary and our conclusions.

\section{General strategy and criteria for dimensional reduction}
\label{DR}

   In this section we set the stage for generalizing the concept
of DR to cases that explicitly involve fermions. First we give a brief
review of screening physics and an overview of how dimensional reduction
comes about: this intuitive guide to our calculation also serves the
purpose of introducing an important concept, the ``mass-shell'' in
screening quantities, that is extensively used throughout the paper.
Then we state the criterion for DR to take place when observables 
whose leading contribution comes from fermions are present, after having 
recapitulated the corresponding criterion for DR in a pure bosonic case. 
Next we discuss in detail the relevant kinematic regions, relevant to
screening physics, where DR could happen both for fundamental fermionic 
fields and for composite operators. Finally, we outline the 
strategy for explicitly verifying whether DR occurs in QCD. Where
appropriate we elucidate similarities and differences with the heavy 
quark expansion.

\subsection{Screening physics and mass-shell condition}
   Finite temperature screening physics can be directly and naturally
formulated in terms of a Lagrangian in the four-dimensional Euclidean space.
Unlike real-time dynamics~\cite{contra}, there is no need to eventually 
analytically continue results to the (3+1)-dimensional Minkowskian space-time, 
since screening phenomena are described by time-independent correlation
functions. Taking advantage of this static nature it is convenient
to Fourier transform the fields in the time direction in terms of
Matsubara frequencies. Then the four-dimensional Euclidean Lagrangian
can be equivalently rewritten as a three-dimensional Euclidean
Lagrangian with an infinite number of Matsubara modes.

   We typically want to study correlations between operators in the
limit of spatial distances much larger than the thermal wavelength, or
more specifically, we only consider correlators at $|\bbox{x}|\gg 1/T$.
This large spatial separation selects a preferred direction, which
we take along the first axis in the four-dimensional Euclidean space
whose zeroth component is the imaginary time. Then the dominant
large-distance contributions to such correlators come from the
lowest singularities in the external momentum variable $p_1$.

   If we consider a weakly interacting theory, it is intuitive that
singularities appear when the external momentum is such that some of the
denominators of the internal propagators vanish. Since the four-momenta are 
Euclidean and we consider massless particles, denominators are of the form
$M^2+\bbox{p}^2$ and vanish either when $M^2=0$ and $\bbox{p}^2=0$ (bosonic
zero mode with vanishing spatial momentum) or when some of the spatial 
components is imaginary, $p_1^2=-M^2$ (bosonic non-zero modes 
$M^2=(2 n\pi T)^2$ and fermionic modes $M^2=((2 n-1)\pi T)^2$). In this
last case $p_1$ is purely imaginary, which corresponds to
an exponentially decaying spatial correlation, a well-expected behavior
when only non-zero modes are involved.

In the complex plane of $p_1$ the condition that the Euclidean
four-momentum be zero becomes $(i p_1)^2-p^2_2-p^2_3=M^2$, which can be
interpreted as the mass-shell condition in a (2+1)-dimensional
Minkowskian space-time for a particle with mass equal to the Matsubara
frequency $M$. The concept of particles being on, close or off their
``mass-shell'' used throughout this paper to describe screening physics
has the precise meaning given by this interpretation of the singularities
in $p_1$ in the original four-dimensional static correlators. It should
be emphasized that the screening singularities we have introduced here 
have no direct relation to real-time dynamics, which corresponds to
singularities in real frequency in the (3+1)-dimensional nonstatic 
correlators~\cite{contra}.

Finally, it is helpful to remind here that thermodynamical quantities
are determined from static Green's functions in the kinematic region
with vanishing or small external momenta, in contrast with the screening
physics directly involving quarks mentioned above.

\subsection{Criteria for dimensional reduction}
\subsubsection{Pure bosonic case}

  DR in the bosonic sector can be explained by the observation that 
there exists a clear scale separation when temperature is high enough. 
Modes with non-vanishing Matsubara frequencies have masses of order $T$, 
while modes with zero Matsubara frequency have masses of order one. 
If one is only interested in the dynamics of the static modes in the 
low momentum regime, nonstatic modes might play negligible role, since their 
effects can only be felt through virtual processes with energies comparable 
to or higher than their masses.

  The above intuitive picture can be formalized within the framework
of perturbation theory in the following way. 
The (D+1)-dimensional Lagrangian ${\cal L}^{D+1}$ is said to undergo DR 
to a specific $D$-dimensional Lagrangian ${\cal L}^D$, in the high-$T$ 
limit and to a given order in the coupling constant, if all the {\it static} 
Greens' functions of ${\cal L}^{D+1}$, $G^{D+1}(\bbox{p},T)$, are equal
to the corresponding Green's functions of ${\cal L}^D$, $G^D(\bbox{p})$, 
up to corrections suppressed by powers of $1/T$:
\begin{equation}
G^{D+1}(\bbox{p},m,T)=G^D(\bbox{p},m)+{\cal O}(|\bbox{p}|/T,m/T)\, ,
\end{equation}
where $m$ represents possible external dimensionful parameters of the theory, 
such as a usual mass parameter. In general, the form and the parameters of
${\cal L}^D$ are determined by the original theory. 

   These naive expectations based on a tree-level power counting
fail because of dynamically generated scales of 
order $T$~\cite{landsman}. Nevertheless, these dynamically generated 
scales only induce corrections proportional to powers of the coupling.
If the coupling is small, the concept is still useful and the reduced 
theory still provides a simplified physics picture.

It is apparent that if the original theory is renormalizable in $D+1$
dimensions, the reduced theory is a super-renormalizable theory in
general, since the coupling in the reduced theory will have positive
dimension in mass units. However, whether the ultraviolet behavior of the
reduced theory is truly improved depends on whether all Matsubara modes,
except a finite number, decouple.
Finally, we remind that beyond tree-level DR manifests explicitly
in the reduced Lagrangian only in certain specific classes of subtraction
schemes~\cite{Appelquist81,landsman}, although the physics of DR for
a given theory is of course a scheme independent phenomenon.
We shall make further comments on this point, when we discuss the
choice of the coupling constant in the reduced theory.

\subsubsection{Fermionic case}

  Due to the antiperiodic boundary condition in the temporal direction,
all fermionic fields have tree-level masses of order $T$. As a consequence,
fermions are usually
treated like implicit degrees of freedom in the high-$T$ limit. However, 
there are situations where we want to study observables that directly 
involve fermions and that are zero if no explicit fermionic dynamics is kept.
A typical example is the correlation between mesonic currents, which
is defined in terms of fermionic fields: the Lagrangian does not include
terms that couple such currents directly to bosonic fields and the leading
contribution is given by fermionic modes. When studying such cases, does
it still make sense to ask whether some of the fermionic modes are more 
important than others? The naive answer might be negative, since there 
exists no obvious scale separation between the lightest fermionic modes
$\pm \pi T$ and the rest.

  This apparent lack of a scale separation is also reflected in
the fact that the typical spatial momenta of the fermionic modes at high 
temperature are not small relative to $T$. 
For example, even in the free theory case, fermions are always 
on their mass-shell and hence $|\bbox{p}|$ is of the order $T$. This 
implies that the expansion in $|\bbox{p}|/T$ is meaningless.

  However, QCD is asymptotically free and quarks are weakly interacting 
in the high-$T$ limit. As previously discussed, in this weakly interacting
regime, quarks are almost on their ``mass-shell''. This intuitive picture
suggests us that the relevant small scale is not the typical spatial
momentum, which is large even in the free theory, but the amount by
which the interaction brings the spatial momentum off the free-theory
mass-shell (the ``off-shellness'') or, in other words, the residual
momentum after the contribution from the mass has been subtracted out.

If we consider, for example, one of the lowest Matsubara modes 
$\omega_{\pm}=\pm \pi T$, we can define the dynamical residual momentum
$\bbox{q}^2\equiv \bbox{p}^2 +(\pi T)^2$ and, similarly to the bosonic
case, we say that the theory undergoes DR, if all the {\it static} 
Greens' functions $G^{D+1}(\bbox{p},T)$ are equal to the corresponding 
Green's functions $G^D(\bbox{p})$ up to corrections suppressed by powers 
of $1/T$:
\begin{equation}
G^{D+1}(\bbox{p},m,T)=G^D(\bbox{p},m)
+{\cal O}(|\bbox{q}|/T,m/T)\, ,
\end{equation}
where again $m$ represents possible external dimensionful parameters 
of the theory. Similarly to bosonic cases, corrections are suppressed
only by powers of the coupling constants instead of powers of $1/T$ when
there are dynamically generated masses proportional to $T$ or, as we
shall see, when the specific dynamics makes the residual momentum 
proportional to $T$.

\subsubsection{Analogy with the heavy-quark expansion}

   The situation we have described for the high-$T$ expansion in
the quark sector shows several similarities with the heavy quark
expansion in QCD~\cite{caswell,lepage,neubert}.
In fact, the heavy quark degrees of freedom can
usually be integrated out leaving power suppressed corrections if
our interest is in light quark observables, such as the $D$ meson.
When we also want to study observables that involve heavy quarks, such
as heavy quarkonia, the explicit heavy degrees of freedom must
be retained. In addition, the effective theory that describes the
heavy quark sector is also derived by expanding in the residual
momentum relative to the heavy quark mass. 

   However, there are two major 
differences between the heavy quark expansion and the high-$T$ expansion
for quarks. One difference is that, in the latter case, we need to 
integrate out an infinite number of Matsubara modes for each flavor,
whereas in the former only the antiquark degree of freedom is
integrated out for each flavor. The other difference is that the coupling 
constant in the high-$T$ reduced theory is, for pure dimensional reasons,
proportional to $T$, while the coupling constant in the heavy-quark 
effective (HQE) theory, is independent of (or at worst logarithmically 
dependent on) the heavy quark mass.  The first difference leads to the
consequence that the HQE theory maintains the original dimensionality
$3+1$ and the high-$T$ effective theory has its dimensionality
reduced by one. The second difference implies that, contrary to the HQE
theory, the accuracy of the high-$T$ effective theory is usually worsened
from powers of $1/T$ to only powers of the coupling.

\subsection{Relevant kinematic region for fermion DR}
  Following the strategy described at the beginning of this section, we
examine here what are the precise kinematic regions where DR can take
place in the fermionic sector for fundamental and composite operators.

   From now on we call light modes the static gluon ($\Omega_n=0$) and
the lightest quarks ($\omega_n=\pm\pi T$), while the rest of the
modes are denoted as the heavy modes.

\subsubsection{Fundamental fields}

  An external light quark close to its mass shell has a momentum of
the form $p=(\omega_\pm,\bbox{p})$ with $\bbox{p}^2\sim -\pi^2T^2$.
For definiteness, let us consider the ``particle'' characterized by
$\omega_+$: the same considerations can be trivially repeated for the
other light mode $\omega_-$. Here and in the following each
quark mode of different Matsubara frequency is regarded as a
``flavor'' in the three-dimensional theory. Quotes are used to 
distinguish this use of the word ``flavor'' from the usual one. Since 
the on-shell condition is defined in Minkowskian space, we choose 
$p_1$ as the ``time'' component or energy in the reduced world (the 
original time component $p_0$ becomes a chirally invariant mass).
Then the residual momentum is defined as $q_1=p_1-i\omega_+$,
$q_2=p_2$ and $q_3=p_3$ with $|q_i|\ll T$. The alternative choice of 
$q_1=p_1+i\omega_+$ can be interpreted as the interchanging of particle 
and antiparticle of the same ``flavor''.

   Of course, one can also choose $p_1\sim i\omega_n T$ with 
$n\ne\pm 1$, i.e. $p_1$ being close to one of the heavy mass-shell.
However, the physically most relevant singularities are those closest
to the origin in the complex plane of $p_1$, since only these are
practically measurable on the lattice. Therefore, we only
limit ourselves to the cases of $p_1\sim i\omega_\pm$.

  A static gluon close to its mass-shell has a momentum of the
form $k=(0,\bbox{k})$ with $|\bbox{k}|\ll T$. Therefore, the
definition of the high-$T$ expansion as an expansion around the
mass-shell of the weakly interacting modes reproduces the usual
condition when applied to the bosonic sector (pure Yang-Mills
case).

\subsubsection{Composite operators}

   The rationale underlying the choice of the relevant kinematic 
regime when composite operators, such as meson and baryon currents,
are present is the same as the one used in the quark and
gluon sector: we expect small deviations from the free theory.

   For a static mesonic current with momentum $k=(0,\bbox{k})$,
the lowest singularity starts at $\bbox{k}^2\sim -(2\pi T)^2$,
as one can easily verify considering the free fermion-bubble graph.
This singularity corresponds to the lowest particle-antiparticle 
threshold, where particles and antiparticles are defined as the
excitations near the two possible mass-shell conditions:
$p_1 = i\omega_+$ and $p_1 = -i\omega_+$.
To be consistent with this condition, the momenta of the particle
and antiparticle in the composite operator have the form
$p=(\omega_+,\bbox{p})$ and $p'=(\omega_+,\bbox{p}')$ with
$p_1\sim i\omega_+$ and $p'_1\sim -i\omega_+$ and $k=p-p'$.
The momenta running along the free fermion-bubble show clearly
why we call, in the reduced theory, quarks with
the same $p_0$ but ``opposite''  $p_1$ components particle and
antiparticle, while we distinguish quarks with different
$p_0$ components using different ``flavor''.

   We could perform a similar analysis for baryonic currents. Since
we do not anticipate any further conceptual problem in this kind
of extension, we consider in this work only mesonic currents.

\subsection{Expansion parameter and power counting}

\subsubsection{Tree level}

  In the relevant kinematic region, we can classify propagators
according to their behavior close to mass-shell and then construct
a corresponding power-counting scheme that characterizes the
behavior of a given graph. For example, it is trivial to verify that,
when close to mass-shell, a light-fermion propagator is of the order 
$T/|\bbox{q}|$, while a static gluon propagator is of the order of 
$T^2/\bbox{k}^2$, relative to propagators of heavy modes.

  The energy-momentum conservation at each vertex forces internal 
quark lines of a graph whose external lines are close to their 
mass-shell to remain themselves close to the mass-shell, unless at
least one of the internal lines is heavy. However, a graph that involves
internal heavy lines is suppressed relative to the same graph with
the heavy lines replaced by light ones, as it is trivially demonstrated
by the propagator classification given above. This result, in turn,
implies that any tree graph with only light external quark or antiquark 
lines is correctly reproduced by the corresponding tree graph in the
reduced theory.

\subsubsection{Loop effects}

   The fact that the dominant contributions to screening observables
come from the kinematic regions where the external line are close to
their mass-shell in the (2+1)-dimensional Minkowskian space implies that 
the relevant power counting that establishes the relative importance
of the difference graphs is not the usual one in Euclidean space. This
necessity of explicitly considering the contributions of the
Minkowskian singularities makes the power counting for graphs
involving loops less straightforward than at the tree level.
Fortunately, the number of graphs at a given order in the loop expansion
is finite. At the one-loop level this number is small. We can just perform
an explicit calculation and isolate the important contributions.

  The strategy we adopt is the following. We consider any graph, 
$G_{4D}(\bbox{p},T)$, in the original theory in 3+1 dimensions, where
$\bbox{p}$ is close to the mass-shell, and we then decompose this graph
into terms recognizable as three-dimensional graphs. We achieve this result
by first separating out those terms in $G_{4D}(\bbox{p},T)$ that involve
only light lines: these terms are by definition reproduced by the
reduced tree-level three-dimensional Lagrangian and we denote them by
$G_{3D}(\bbox{p},T)$. The remaining contributions to
$G_{4D}(\bbox{p},T)$ involve at least one heavy line and we define them
as $G^H(\bbox{p},T)$. 
We say that DR occurs if all such $G^H(\bbox{p},T)$'s can be made local,
i.e. if they result in a polynomial in the residual momentum $\bbox{q}$.
In other words, DR occurs, if
\begin{equation}
G_{4D}(\bbox{p},T)=G_{3D}(\bbox{p},T)+G^H(\bbox{p},T)
=G_{3D}(\bbox{p},T)+G^H(0,T)+{\cal O}(|\bbox{q}|/T)\, ,
\end{equation}
where $G^H(0,T)$ has either the form of terms already present in the
tree-level reduced Lagrangian (its effect is the renormalization of the
relevant parameters) or the form of a new renormalizable vertex. More
generally, the reduced theory needs also to contain nonrenormalizable 
vertices if we want to reproduce graphs to higher orders; these terms
can be treated consistently in the context of an effective theory
(see Sec.~\ref{NRred}).

\section{The Reduced Theory: Calculation} 
\label{DRcalc}
\subsection{The tree-level Lagrangian}
\label{redtree}

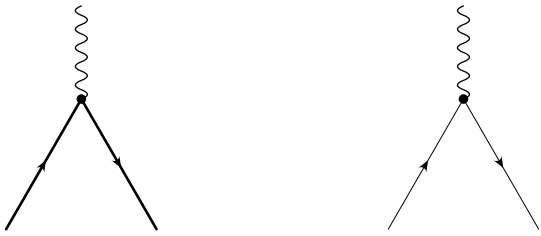
\begin{figure}
\vspace{0.1cm}
\caption[]{Vertices of ${\cal L}^0_{\text{RD}}$ that involve only
light modes. A wiggly line represents a static gluon, while a thick
(thin) solid line represents a quark of frequency $\omega_+$
($\omega_-$), respectively.}
\label{fig1}
\end{figure}

At the tree level the reduced theory can be simply obtained by
Fourier transforming the QCD Lagrangian and then retaining only
the static gluonic fields and the quark fields (after rescaling
a factor of $\sqrt{T}$) with lowest Matsubara
frequencies $(\omega_\pm=\pm\pi T$), 
\begin{equation}
{\cal L}^0_{\text{RD}}=-{1\over 4}F^a_{ij} F^a_{ij}
-{1\over 2}(D_i A_0)^a(D_i A_0)^a +\sum_{l=\pm}\bar{\psi}_l
\Bigl[-\Bigl(\omega_l+g_3 A_0^a{\lambda^a\over 2}\Bigr)\gamma_0
+i\gamma_i D_i\Bigr]\psi_l \, ,
\label{qcd_3d}
\end{equation}
where
$F_{ij}^a=\partial_i A_j^a-\partial_j A_i^a-g_3 f^{abc} A_i^b A_j^c$,
$(D_i A_0)^a=\partial_i A_0^a-g_3 f^{abc} A_i^b A_0^c$ and
$D_i\psi=\partial_i\psi-i g_3 A_i^{a}(\lambda^a/2)\psi$.
For simplicity we have assumed that quarks are massless and have
$N_f$ flavors. 
Figure~\ref{fig1} shows the two vertices of ${\cal L}^0_{\text{DR}}$
that involve quarks: the graphical notation is such that a thick (thin) 
solid line represents a quark with frequency $\omega_+$ ($\omega_-$) and 
a wiggly line represents a static gluon. 

    The coupling $g_3$ is related  to the four-dimensional coupling 
through $g_3^2=g^2(\mu)T$. At the tree level the subtraction scale 
$\mu$ is not yet specified. Since ${\cal L}^0_{\text{RD}}$ is a 
super-renormalizable theory in three dimensions, all the dynamical 
scales must be set by the coupling constant $g_3^2=g^2(\mu)T$ and 
temperature $T$.

  Of course, once loop corrections are included, the reduced theory
in Eq.~(\ref{qcd_3d}) acquires new vertices and the coupling
constant $g_3^2$ has a more complicated  dependence on the original 
coupling $g^2(\mu)$. For example, $g_3^2$ can receive corrections
such as $g^4(\mu)T$ and so on. However, since QCD is asymptotically
free, the appropriately defined coupling constant (DR is only
manifest in subtraction schemes that require
$\mu\sim T$~\cite{landsman,lambda,drl}) has the asymptotic behavior
$g^2(\mu\sim T)\sim 1/\ln T$. Therefore, the corrections to the
tree-level form of $g_3$ should not modify the fact that the two
dynamical scales be $g_3^2\approx g^2(\mu\sim T)T$ and $T$ itself at
high temperature. The criterion according to which we choose $\mu\sim T$ 
is in general to minimize loop corrections to the leading result. We
shall discuss the precise choice of the proportionality constant later
in the paper.

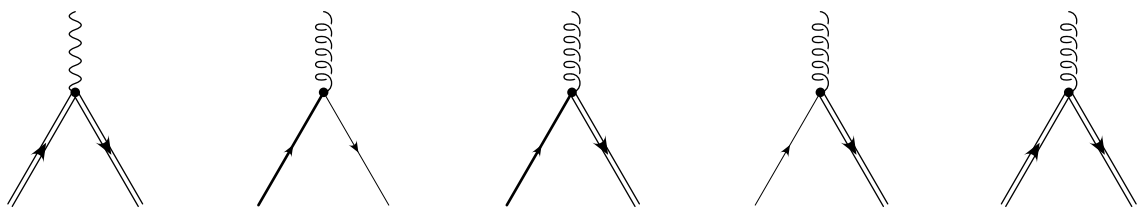
\begin{figure}
\vspace{0.1cm}
\caption[]{Vertices of the original Lagrangian that are not present in
${\cal L}^0_{\text{RD}}$, since they involve at least one heavy mode
($\Delta{\cal L}_{\text{H}}$). A double line
denotes a quark of frequency $|\omega_n|\ge 3\pi T$, while
a spring-like line denotes a nonstatic gluon mode. The first vertex
is ``flavor-conserving'' and the rest are all ``flavor-changing''.}
\label{fig2}
\end{figure}

   The vertices that involve light quarks in the reduced theory,
the ones shown in Fig.~\ref{fig1}, are not the only vertices present 
in the original theory. There are additional vertices that involve
at least one heavy mode; we show them in Fig.~\ref{fig2} with the
notation that a double line stands for a heavy quark mode and a
spring-like line stands for a nonstatic gluon mode. These
vertices are collectively called $\Delta{\cal L}_{\text{H}}$.
In the first vertex the fermionic line does not change its ``flavor'',
since the gluon is static, and we call it ``flavor-conserving''. All the
other vertices in Fig.~\ref{fig2} involve a nonstatic gluon and, therefore,
the fermionic line changes its ``flavor''; we call these vertices
``flavor-changing''.

Since we are only interested in graphs with light modes in the external 
lines, energy-momentum conservation implies that these heavy
vertices cannot contribute at the tree level. Our job is to verify
whether the corrections induced by these heavy vertices at the one-loop 
order can be accounted for, in the high-$T$ limit,  either by readjusting 
the parameters of ${\cal L}^0_{\text{DR}}$ or by adding additional local 
and renormalizable vertices.

\subsection{One-loop graphs}
\label{onepi}

  In the following calculations we use the dimensional regularization 
for spatial integrals,
\begin{equation}
\int{d^4 k\over(2\pi)^4}\rightarrow
T\sum_{n=-\infty}^\infty \, \mu^{2\epsilon}
\int{d^{3-2\epsilon}\bbox{k}\over(2\pi)^{3-2\epsilon}}
\equiv T\sum_{n=-\infty}^\infty\int [d\bbox{k}]\, ,
\label{dim_reg}
\end{equation}
and the $\overline{\text{MS}}$ subtraction scheme.
For convenience, we work in the Feynman gauge and use the
Euclidean Feynman rules given by Ramond~\cite{ramond}. 

The  subtraction  point $\mu$ needs to be proportional to $T$ to 
make the DR manifest~\cite{landsman}, then we define the
$T$-dependent coupling constant 
\begin{equation}
\tilde{g}^2(T)\equiv g_{\overline{\text{MS}}}^2(\mu)
\label{coupling}
\end{equation}
by choosing
\begin{equation}
\mu=\Bigl(4\pi\, e^{-\gamma_E-c}\Bigr)\, T \, .
\label{scale_mu}
\end{equation}
The temperature independent constant $c$ can be read as a convenient way
of parameterizing one of the possible subtraction schemes. The specific
choice $c=(N/2-2N_f\ln4)/(11N-2N_f)$ corresponds, for instance, to the
scheme that makes DR optimal for the pure-gluon effective action in the
background field method~\cite{lambda}. We shall comment later on
alternative choices such as the one that makes DR optimal in terms of 
the quark-gluon vertex function.
In the following we also use the auxiliary coupling
\begin{equation}
{\cal G}^2(T)\equiv \tilde{g}^2(T)/(16\pi^2) \, ,
\label{newcoupling}
\end{equation}
for the sake of making formulae more compact.

   Without any loss of generality, we only show results for cases with
external quark-line frequency   $\omega=\omega_+\equiv\pi T$; trivial 
modifications yield the corresponding results for the other light 
``flavor'' with $\omega=\omega_-\equiv-\pi T$. Moreover, we select the 
particle sector,
which we have defined as the kinematic region where the first component
of the spatial momentum is close to $i\omega_+$; again there are only
trivial  differences for the antiparticle sector, which is defined
by the alternative choice of the spatial component being close 
to $-i\omega_+$, and that is separated from the particle sector by 
a off-(mass)shellness $2\pi T$.

  In the Matsubara-frequency loop-sum, the term with zero frequency
can be easily recognized as the contribution from the three-dimensional
tree-level Lagrangian itself, up to trivial factors of $T$. Therefore,
when computing the loop corrections due to the heavy Matsubara modes,
we leave out the term with $n=0$, which is the direct
contribution from the light modes and it is already generated by
the reduced theory.

   The calculation of the one-loop amplitudes and their high temperature 
expansion around the appropriate kinematic regions can be done following
a standard procedure, which, in general, involves the following steps:\\
(1) combine the denominators by means of the Feynman parameter 
representation and perform the spatial momentum integral;\\
(2) shift the external momenta: $\bbox{p}=\bbox{q}+i\omega_+\hat{e}_1$ and 
$\bbox{p}'=\bbox{q}'+i\omega_+\hat{e}_1$; \\
(3) expand the result in terms of the residual momenta over $T$, 
    e.g. $|\bbox{q}|/T$;\\
(4) perform the integrals over the Feynman parameters;\\
(5) perform the Matsubara sum and express the result in terms of the 
    Riemann zeta function.\\
In the following we only give the final results obtained by the application
of the above-described procedure.

\subsubsection{Quark self-energy}

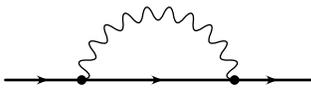
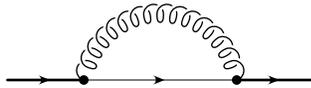
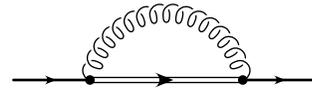
\begin{figure}
\vspace{0.1cm}
\caption[]{Feynman graphs for the quark self-energy correction.
Graph (a) is generated by ${\cal L}^0_{\text{RD}}$, while
graphs (b) and (c) involve ``flavor-changing'' vertices from 
$\Delta{\cal L}_{\text{H}}$. The external quark carries four-momentum
$p=(\omega_+,\bbox{p})=(\omega_+,i\omega_+ +q_1,q_2,q_3)$.}
\label{fig3}
\end{figure}

  The quark self-energy correction due to heavy modes is found by 
calculating the graphs drawn in Figs.~\ref{fig3} (b) and (c) with 
external momentum $p=(\omega_+,\bbox{p})=(\omega_+,i\omega_+ +q_1,q_2,q_3)$.
The result is:
\begin{equation}
i\Sigma(p)=-C_f \, {\cal G}^2(T)\left\{\Bigl[
(X+2c)\, p\cdot\gamma+X_{\mu\nu}p_\mu\gamma_\nu\Bigr]
+{\cal O}\Bigl({|\bbox{q}|\over T}\Bigr)\right\}\, ,
\end{equation}
where $C_f=(N^2-1)/(2N)$ and the coefficients $X$ and $X_{\mu\nu}$ 
are pure numbers that are listed in Table~\ref{table1}. In this
Table there appear the derivative of the Riemann zeta function
evaluated at $-1$, $\zeta'(-1)$, and the Euler's constant gamma
$\gamma_E$, whose approximate numerical values are
$\zeta'(-1)\approx -0.16542$ and $\gamma_E\approx 0.57722$.

We find that the heavy mode correction to the self-energy $\Sigma(p)$ 
is suppressed relative to the tree-level piece $ip\cdot\gamma$ by the 
factor ${\cal G}^2(T)$. In addition, we also point out that, as expected,
no chiral-symmetry-breaking mass-term for the quark self-energy has
been perturbatively generated, in spite of the fact that other
noncovariant terms have instead appeared.

\subsubsection{Quark-gluon vertex}

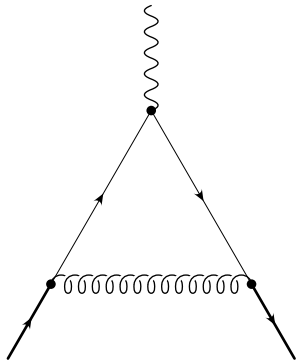
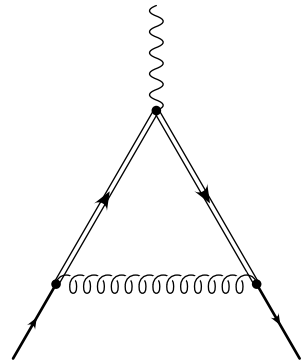
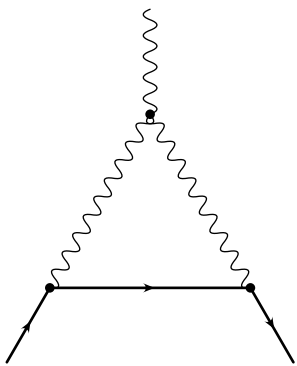
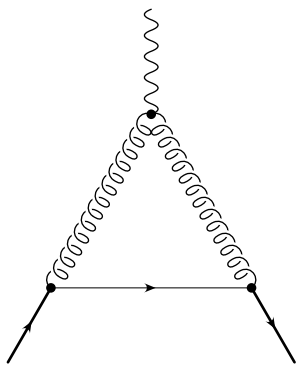
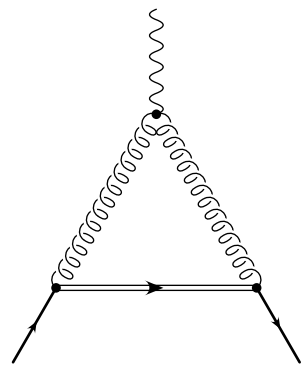
\begin{figure}
\vspace{0.1cm}
\caption[]{Feynman graphs for the quark-gluon vertex correction.
Graphs (a) and (d) are generated by ${\cal L}^0_{\text{RD}}$,
while graphs (b), (c), (e) and (f) involve ``flavor-changing''
vertices from $\Delta{\cal L}_{\text{H}}$.
The incoming quark carries four-momentum 
$p=(\omega_+,\bbox{p})=(\omega_+,i\omega_+ +q_1,q_2,q_3)$,
and the outgoing antiquark carries four-momentum
$p'=(\omega_+,\bbox{p}')=(\omega_+,i\omega_+ +q'_1,q'_2,q'_3)$,
with $|\bbox{q}|\ll T$ and $|\bbox{q}'|\ll T$.}
\label{fig4}
\end{figure}

  The corrections of the heavy modes to the quark-gluon vertex come
from two types of graphs: graphs that have an analogue in QED, 
Figs.~\ref{fig4} (b) and (c), and graphs that are intrinsically 
nonabelian, Figs.~\ref{fig4} (e) and (f). In the following the
momentum of the incoming quark is labeled by
$p=(\omega_+,\bbox{p})=(\omega_+,i\omega_+ +q_1,q_2,q_3)$ and
the momentum of the outgoing quark by
$p'=(\omega_+,\bbox{p}')=(\omega_+,i\omega_+ +q'_1,q'_2,q'_3)$.

The first type of graphs yields:
\begin{equation}
\Gamma_\mu(p,p')=i\tilde{g}(T){\lambda^a\over 2}
\Bigl(C_f-{C_{ad}\over 2}\Bigr) {\cal G}^2(T)\left\{
\Bigl[(Y+2c)\, \gamma_\mu+ Y_{\mu\nu}\gamma_\nu\Bigr]
+{\cal O}\Bigl({|\bbox{q}|\over T},{|\bbox{q}'|\over T}\Bigr)
\right\}\, ,
\end{equation}
where $C_{ad}=N$ and the coefficients $Y$ and $Y_{\mu\nu}$ are given 
in Table~\ref{table1}.

  The second type of graphs yields:
\begin{equation}
\Gamma'_\mu(p,p')=i\tilde{g}(T){\lambda^a\over 2}
C_{ad}\,{\cal G}^2(T)\left\{
\Bigl[(Z+3c)\, \gamma_\mu+ Z_{\mu\nu}\gamma_\nu\Bigr]
+{\cal O}\Bigl({|\bbox{q}|\over T},{|\bbox{q}'|\over T}\Bigr)
\right\}\, ,
\end{equation}
and again the coefficients $Z$ and $Z_{\mu\nu}$ are listed in 
Table~\ref{table1}.

  In Table~\ref{table1} one can notice that there exist precise relations
between some of the entries for the quark self-energy ($X$'s) and of the 
abelian part of the quark-gluon vertex ($Y$'s): the reason of these 
relations is the existence of generalized QED-like Ward identities at 
finite temperature, due to the static gauge invariance. The fact that 
our results verify these Ward identities serves as a very useful 
consistency check.

  We find again that the heavy mode corrections to the tree-level coupling
coming from $\Gamma_\mu(p,p')$ and $\Gamma'_\mu(p,p')$ are suppressed 
relative to the tree-level piece $-i\tilde{g}(T)\gamma_\mu$ by the 
factor ${\cal G}^2(T)$. In addition, there are also new vertices not
present at $T=0$, but these vertices are also order of ${\cal G}^2(T)$
relative to $-i\tilde{g}(T)\gamma_\mu$.

\subsubsection{Vacuum polarization}
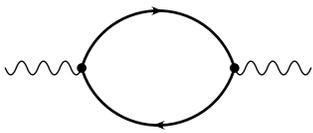
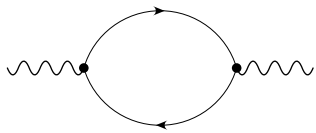
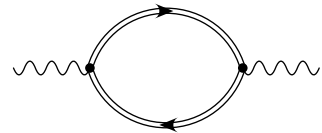
\begin{figure}
\vspace{0.1cm}
\caption[]{Feynman graphs for the quark contributions to the vacuum 
polarization tensor. 
Graphs (a) and (b) are generated by ${\cal L}^0_{\text{RD}}$,
while graph (c) involves ``flavor-conserving'' vertices from 
$\Delta{\cal L}_{\text{H}}$. The external gluon carries four-momentum
$k=(0,\bbox{k})$, with $|\bbox{k}|\ll T$.}
\label{fig5}
\end{figure}

The contributions to the vacuum polarization tensor coming from both
light and heavy quarks are given by the graphs shown in
Figs.~\ref{fig5} (a), (b) and (c). There are a few small differences 
relative to the previous two cases. One is that we can consider the
three graphs together and we do not need to separate out the contribution
from the lowest modes, since all three are infrared finite. In
addition, the external gluon line is static $k=(0,\bbox{k})$ and the
loop Matsubara sum is now fermionic $p=(\omega_n,\bbox{p})$,
with $\omega_n=(2n-1)\pi T$.

In the magnetic sector $\Pi_{ij}(\bbox{k})$ remains transverse
\begin{equation}
\Pi_{ij}(\bbox{k})=(k_i k_j-\bbox{k}^2\delta_{ij})\delta_{ab}N_f
{g^2(\mu)\over 24\pi^2}\biggl\{{1\over\epsilon}+\gamma_E
+\ln\Bigl({4\mu^2\over\pi T^2}\Bigr)
+{\cal O}\Bigl({\bbox{k}^2\over T^2}\Bigr)\biggr\}\, .
\end{equation}
After renormalizing at the scale $\mu=4\pi e^{-\gamma-c}T$, we see 
that the only effect of $\Pi_{ij}$ is to give a finite wave-function 
renormalization to the static gluon compared to the zero temperature
case.

  We also find that $\Pi_{i0}=\Pi_{0i}=0$, which in turn implies that 
there is no mixing between electric and magnetic components in the
static sector induced by the nonstatic modes.

   The result for the electric part is
\begin{equation}
\Pi_{00}(\bbox{k})=-\delta_{ab}N_f {g^2(\mu)\over 24\pi^2}
\Biggl\{4\pi^2T^2+\bbox{k}^2\biggl[{1\over\epsilon}
+\gamma_E+\ln\Bigl({4\mu^2\over\pi T^2}\Bigr)-1\biggr]
+{\cal O}\Bigl({\bbox{k}^2\over T^2}\Bigr)\Biggr\}\, .
\end{equation}
After the wave-function renormalization, which again only receives a
finite contribution with respect to the zero temperature case, we are
still left with an additional $T^2$ term that implies a mass generation 
for the time component of the static gluon field, which is the well-known
analogue of the Debye screening in QED.

\subsection{Summary of one-particle irreducible graphs}
  
  The effect of the heavy modes of the original four-dimensional Lagrangian,
which are no longer present in the reduced theory, can be reproduced,
at the one-loop level and in the relevant kinematic region, by the
following three types of corrections to the tree-level (2+1)-dimensional
reduced Lagrangian.\\
(a) The quark bubble insertion corrects the static gluon propagator
    and is suppressed by a factor ${\cal G}^2(T)$. It also generate a
    mass term of order ${\cal G}^2(T)T^2$ for $A_0$.\\
(b) The quark self-energy insertion corrects the lightest quark
     or antiquark propagator and, in addition, generates new terms that
     are not present at zero temperature. Corrections and new terms are 
     all suppressed by a factor ${\cal G}^2(T)$.\\
(c) The static gluon-quark vertex insertion corrects the tree-level vertex
    and generate new vertices. Again corrections and additional vertices
    are suppressed by a factor ${\cal G}^2(T)$.\\
(d) In this section we only considered graphs explicitly involving quark
fields. Since the on-shell condition for bosonic fields are the same with
or without the presence of fermions, graphs containing purely gluons or
ghosts have already been considered in Refs.~\cite{landsman,kajantie}.

   In summary, we find two kinds of corrections in the infrared limit, 
namely when the light modes are close to their mass-shell.
There are corrections that are directly generated by the corresponding 
one-loop graphs in the reduced theory, once the coupling constant is 
properly chosen. In addition, there are also one-loop corrections that
are not contained in the tree-level reduced theory. However, all these
terms are infrared finite to the order considered and are suppressed by
a factor ${\cal G}^2(T)$ and hence subleading. The fact that this
corrections are infrared finite implies that they can be accounted for
by adding new local and renormalizable vertices to the reduced theory. 

\section{Composite operators}
\label{CO}

In confined theories, such as QCD, the fundamental degrees of freedom
are not manifest in the spectrum and it is necessary to use
composite operators to probe physical particles of the theory.
For example, one uses composite operators as interpolating fields for
mesonic and baryonic states. Therefore, the study of how composite 
operators are reproduced in the reduced theory is necessary to have
a complete picture of the DR physics. We shall find that considering
composite operators introduces new features that are not trivial
extensions of what already discussed.

For the sake of concreteness, we focus our attention on flavor
nonsinglet mesonic currents. Generalization to other cases,
such as baryonic operators, can be done analogously. 
At the tree level, the static limit of
these currents can be written as a sum over Matsubara modes
\begin{equation}
{\cal O}_\Gamma\equiv
\int\!\! d\tau\,\,\bar{\psi}(\tau,\bbox{x})\Gamma\psi(\tau,\bbox{x})
\propto \sum_n\bar{\psi}_n(\bbox{x})\Gamma\psi_n(\bbox{x})\, ,
\end{equation}
where $\Gamma$ is any of the sixteen Dirac matrices.
As discussed in Sec.~\ref{DR}, the kinematic region of interest
is the one where the lightest modes
are close to their mass-shell: $2\pi T$ in mesonic and $3\pi T$ 
in baryonic cases, respectively. Then at the tree level the high
temperature limit implies that the dominant contribution comes from
the operator obtained by using only the lowest Matsubara quark modes.

\subsection{One-loop correction}
  The procedure for calculating the one-loop correction to composite
operators is similar to the one used for calculating the one-loop 
vertex correction,
except that the momentum carried by the composite operator is now
close to the particle-antiparticle mass-shell, in contrast with
the momentum carried by a static gluon in the vertex correction.
The explicit graphs are shown in Figs.~\ref{fig6} (a), (b) and (c).
The choice of the flavor nonsinglet current avoids the
mixing with gluonic fields\footnote{States that mix with gluonic fields
are dominated by the bosonic zero modes and can be described by the pure
glue reduced theory with the addition of local operators that couple
quarks and gluonic states.}.

More specifically, the external quark and antiquark momenta are
expanded according to $\bbox{p}=\bbox{q}+i\omega_+\hat{e}_1$
and $\bbox{p}'=\bbox{q}'-i\omega_+\hat{e}_1$, where $\bbox{q}$ and
$\bbox{q}'$ are supposed to be small relative to $T$. This choice
reproduces the expected on-shell condition for mesonic currents 
$\bbox{p}-\bbox{p}'\approx 2i\omega_+\hat{e}_1$. 
Since this kinematic difference implies a new feature, we give 
more details in this case.

The one-loop correction to the composite operator is proportional to 
\begin{equation}
\Delta{\cal O}_\Gamma=g^2(\mu)\,C_f\,V_{\mu\nu}(p,p')\,
\gamma_\alpha\gamma_\mu\Gamma\gamma_\nu\gamma_\alpha\, ,
\end{equation}
where $V_{\mu\nu}(p,p')$ is defined as
\begin{equation}
V_{\mu\nu}(p,p')\equiv 2T\sum_{n\ne 0}\int[d\bbox{k}]
\int_0^1d\alpha\int_0^{1-\alpha}d\beta\,
{(p+k)_\mu(p'+k)_\nu\over\Bigl[(1-\alpha-\beta) k^2
+\alpha (p+k)^2+\beta (p'+k)^2\Bigr]^3} \, .
\label{V_def}
\end{equation}
This definition obviously implies that
$V_{\mu\nu}(p,p')=V_{\nu\mu}(p',p)$.
The term in the Matsubara sum with $n=0$, which is represented
in Fig.~\ref{fig6} (a), has been excluded from $V_{\mu\nu}$,
since this term is directly reproduced by the reduced theory
${\cal L}^0_{\text{RD}}$, given by Eq.~(\ref{qcd_3d}).

Fig.~\ref{fig6} (c), i.e. the terms with $n\neq -1, 0$, can be
checked to be infrared finite; their explicit contribution
to $V_{\mu\nu}(p,p')$ is, up to corrections
of ${\cal O}(|\bbox{q}|/T,|\bbox{q}|/T)$,
\begin{mathletters}
\begin{eqnarray}
V_{00}^{(c)}(p,p')&=&{1\over 64\pi^2}\biggl\{
{1\over\epsilon}+\ln\Bigl({\mu^2\over 4\pi T^2}\Bigr)
+\gamma_E+ W_{00}\biggr\}\, , \\
V_{ij}^{(c)}(p,p')&=&{\delta_{ij}\over 64\pi^2}\biggl\{
{1\over\epsilon}+\ln\Bigl({\mu^2\over 4\pi T^2}\Bigr)
-\gamma_E\biggr\}
+{\delta_{i1}\delta_{j1}\over 64\pi^2}
 W_{11}\, , \\
V_{01}^{(c)}(p,p')&=&-V_{10}^{(c)}(p,p')=
-{i\over 64\pi^2}W_{01}\, , \\
V_{0i}^{(c)}(p,p')&=&-V_{i0}^{(c)}(p,p')=0\, 
\,\,\,\,\,{\text{for}}\,\,i=2,3\, ,
\end{eqnarray}
\end{mathletters}
where the numerical coefficients $W_{00}$, $W_{11}$, and $W_{01}$
are given in Table~\ref{table1}. After taking care of the
$1/\epsilon$ terms, which have exactly the same form as at $T=0$,
using the standard composite operator renormalization, there only 
remain the infrared finite terms that can be reproduced
by correcting the tree-level current ${\cal O}_\Gamma$ in the 
reduced theory with the addition of local operators.

\subsection{Additional infrared singularity}
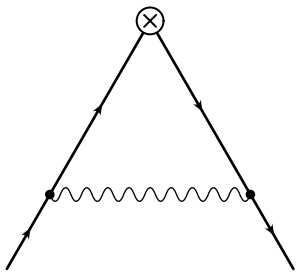
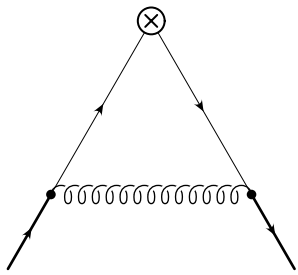
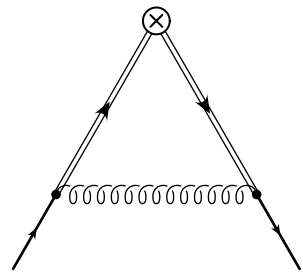
\begin{figure}
\vspace{0.1cm}
\caption[]{Feynman graphs for the one-loop correction to the 
composite operator. Graph (a) is generated by ${\cal L}^0_{\text{RD}}$,
while graphs (b) and (c) involve ``flavor-changing'' vertices from
$\Delta{\cal L}_{\text{H}}$. The incoming quark carries the four-momentum 
$p=(\omega_+,\bbox{p})=(\omega_+,i\omega_+ +q_1,q_2,q_3)$,
and the outgoing antiquark carries four-momentum
$p'=(\omega_+,\bbox{p}')=(\omega_+,-i\omega_+ +q'_1,q'_2,q'_3)$,
with $|\bbox{q}|\ll T$ and $|\bbox{q}'|\ll T$.}
\label{fig6}
\end{figure}

  Now let us focus on the remaining term shown in Fig.~\ref{fig6} (b),
corresponding to $n=-1$ in $V_{\mu\nu}(p,p')$ of Eq.~(\ref{V_def}).
In this term a heavy gluon
of frequency $\Omega_n=-2\pi T$ is exchanged with a corresponding
change of the ``flavor'' (frequency) of the quarks at the vertices.
The expansion around the mass-shell of this term yields, apart from
terms of ${\cal O}(|\bbox{q}|/T,|\bbox{q}'|/T)$,
\begin{eqnarray}
& &V_{\mu\nu}^{(b)}(p,p')= {1\over 16\pi^2} \times \\
& &\quad\quad\int_0^1\!\! d\alpha\int_0^{1-\alpha}\!\!\!\! d\beta\,
{\delta_{\mu 0}\delta_{\nu 0}
+i\delta_{\mu 0}\delta_{\nu 1}(1-\alpha+\beta)
-i\delta_{\mu 1}\delta_{\nu 0}(1+\alpha-\beta)
+\delta_{\mu 1}\delta_{\nu 1}[1-(\alpha-\beta)^2]
\over \bigl[(2-\alpha-\beta)^2+2a\alpha(1-\alpha)
+2b\beta(1-\beta)-2\alpha\beta(2-a-b)\bigr]^{3/2}}\, , \nonumber
\end{eqnarray}
where $a\equiv iq_1/(\pi T)$ and $b=-iq'_1/(\pi T)$. The 
Feynman-parameter integrals can be carried out with the result
\begin{eqnarray}
V_{\mu\nu}^{(b)}(p,p')=&-&{\ln(a+b)\over 64\pi^2}
\Bigl[\delta_{\mu 0}\delta_{\nu 0}
+i\delta_{\mu 0}\delta_{\nu 1}-i\delta_{\mu 1}\delta_{\nu 0}
+\delta_{\mu 1}\delta_{\nu 1}\Bigr] \nonumber \\
&+&{1\over 32\pi^2}
\Bigl[\delta_{ij}+\delta_{i1}\delta_{j1}(4\ln2-3)\Bigr]
+{\cal O}\Bigl({\bbox{q}|\over T},{|\bbox{q}'|\over T}\Bigr)\, ,
\label{log_div}
\end{eqnarray}
which is clearly logarithmically divergent in the infrared.

    The situation is summarized as follows.\\
(1) The term represented by the graph in Fig.~\ref{fig6} (a) diverges 
linearly and gives the leading infrared contribution coming from the 
composite operator: this term only involves the lowest Matsubara modes 
and, therefore, its infrared physics is exactly reproduced by the 
reduced theory ${\cal L}^0_{\text{DR}}$. \\
(2) The terms represented by the graph in Fig.~\ref{fig6} (c) are 
infrared finite: these terms involve heavy Matsubara modes and their 
contribution is not generated by the tree-level reduced Lagrangian 
${\cal L}^0_{\text{RD}}$, but they can be compensated by adding
new local operators to ${\cal O}_\Gamma$.\\
(3) The term represented by the graph in Fig.~\ref{fig6} (b) diverges
logarithmically and gives a subleading infrared contribution compared
to graph (a): this term involves the lowest Matsubara modes for the
quark lines and a heavy mode for the gluon line and, therefore, it is 
not present in the reduced theory. In addition, this infrared
logarithmic behavior apparently implies that it cannot be generated
by adding a local correction to the operator. In other words, we need to
settle the question whether it is possible to add to the tree-level operator 
in the reduced theory  higher-dimensional operators that might generate 
the same logarithmic singularity. This option is easily ruled out, once
one recognizes that this logarithmic singularity is associated with the 
total external momentum carried by the composite current $q_1-q'_1$ 
and is related to the particle-antiparticle (of the ``$-$'' mode)
production threshold, as can be seen explicitly from Fig.~\ref{fig6} (b).

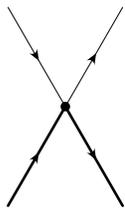
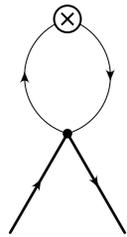
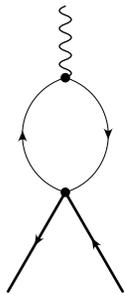
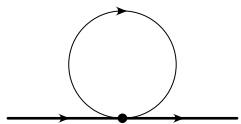
\begin{figure}
\vspace{0.1cm}
\caption[]{Four-quark vertex (a) and its contribution to (b) one-loop
correction of composite operators, (c) one-loop correction to the
fundamental vertex and (d) one-loop correction to the quark self-energy.}
\label{fig7}
\end{figure}

  We are left with the option of adding a higher-dimensional vertex 
to the reduced theory. We can easily guess the form of this new vertex
from the fact that the logarithmic singularity comes from the kinematic
region of  Fig.~\ref{fig6} (b) where the particle and antiparticle are 
close to their mass-shell, while the heavy gluon is far off its mass-shell,
i.e. the gluon line ``contracts'' to a point. In fact, integrating
out the heavy gluon yields the four-quark vertex
\begin{equation}
\Delta{\cal L}_F={\tilde{g}^2(T)\over 4\pi^2T}
\Bigl(\bar{\psi}_+\gamma_\mu{\lambda^a\over 2}\psi_-\Bigr)
\,\Bigl(\bar{\psi}_-\gamma_\mu{\lambda^a\over 2}\psi_+\Bigr) \, ,
\label{fourquark}
\end{equation}
which is depicted in Fig.~\ref{fig7} (a). If we add this four-quark 
vertex to the reduced Lagrangian, there appears a new contribution,
shown in Fig.~\ref{fig7} (b), to the composite operator one-loop 
calculation. This new contribution gives the same logarithmic singularity 
of Eq.~(\ref{log_div})
\begin{eqnarray}
\Delta V_{\mu\nu}(p,p')&=&-{\ln(a+b)\over 64\pi^2}
\Bigl[\delta_{\mu 0}\delta_{\nu 0}
+i\delta_{\mu 0}\delta_{\nu 1}-i\delta_{\mu 1}\delta_{\nu 0}
+\delta_{\mu 1}\delta_{\nu 1}\Bigr]\nonumber \\
&+&{1\over 32\pi^2}\Bigl[(\delta_{\mu 0}\delta_{\nu 0}
+i\delta_{\mu 0}\delta_{\nu 1}-i\delta_{\mu 1}\delta_{\nu 0}
+\delta_{\mu 1}\delta_{\nu 1})\ln 2
-(\delta_{i1}\delta_{j1}+\delta_{ij})/2\Bigr]\, ,
\end{eqnarray}
while the finite differences can now be compensated by local
corrections to the operator.

  Of course, if we add this new four-quark vertex to the reduced
Lagrangian, we need also to consider its additional contribution to 
the fundamental one-particle irreducible graphs which have already 
been calculated in Sec.~\ref{DRcalc}. In particular, we must check 
that the corrections to those
graphs induced by this four-quark term are infrared finite.
At the one-loop level, we need to consider only two graphs: the one
depicted in Fig.~\ref{fig7} (c), which contributes to the fundamental 
vertex, and the one depicted in Fig.~\ref{fig7} (d), which contributes
to the quark self-energy. An explicit calculation shows that these two
contributions are indeed infrared finite. Explicitly, the
correction to the vertex, Fig.~\ref{fig7} (c), is
\begin{equation}
\tilde{\Gamma}_\mu(p,p')=-i\tilde{g}(T){\lambda^a\over 2}
\Bigl(C_f-{C_{ad}\over 2}\Bigr)\,
{\cal G}^2(T)\left\{\gamma_0\gamma_\mu\gamma_0
+{\cal O}\Bigl({|\bbox{q}|\over T},{|\bbox{q}'|\over T}\Bigr)
\right\}\, ,
\end{equation}
while the correction to the self-energy, Fig.~\ref{fig7} (d), contributes
to the chirally invariant mass:
\begin{equation}
i\tilde{\Sigma}(p)= C_f \,{\cal G}^2(T)\,
\Bigl[2\omega_+\gamma_0\Bigr]\, .
\end{equation}

  Finally, one might worry that the four-quark term given in
Eq.~(\ref{fourquark}), since its mass dimension is four, might
be nonrenormalizable in the reduced theory, at least by a naive 
power-counting argument. However, the use of dimensional regularization 
makes all graphs shown in Figs.~\ref{fig7} ultraviolet finite.
In fact, it is the very fact that this operator has mass dimension
four that makes its contribution, which is suppressed by one power
of $1/T$ at the tree level ($\Delta{\cal L}_F\sim 1/T$),
be actually only suppressed by $\tilde{g}^2(T)$ in the one-loop
graphs of Figs.~\ref{fig7} (b), (c) and (d).

\subsection{Comments to the one-loop calculation}

   In principle, one could go on and examine higher orders in the loop
expansion. However, this additional effort is useless in this 
framework. 
In fact, the naive expansion in powers of $1/T$ is only possible, in
general, to the leading order in $\tilde{g}^2(T)$. First, thermal 
masses are eventually  generated at some order in the coupling and 
these masses break the implicit assumption that all the relevant 
momenta can be made small at will. 
Second, the power counting argument that is discussed in the next 
section shows that the residual momenta (off-shellness) in
quantities such as screening masses is of the order
$q^2\sim\tilde{g}^2(T)T^2$, which implies that terms proportional to
$q^2/T^2$ are not suppressed by power of $1/T$ but only by powers
of $\tilde{g}^2(T)$, which vanishes only logarithmically.

At this point we could write down the reduced theory at the one-loop
level, utilizing the calculations done earlier in the last and this
sections. However, one important point about the expansion of the
off-shellness is the following. When we choose the first component of
the momentum $(p_1)$ to be close to either $\pm i \pi T$ as the starting
point of the off mass-shell expansion, this choice breaks the symmetry
between what we call particle and antiparticle in the reduced theory.
If we insist that the reduced theory be formally relativistic,
this asymmetry results in the entanglement of corrections of different
orders in $\tilde{g}^2(T)$ at a given order in loop expansion.
In other words, the heavy quarks of the (2+1)-dimensional reduced 
theory become nonrelativistic (NR) in the high-$T$ limit,
as we shall see in the next section, and the relativistic formalism
has the effect of retaining degrees of freedom that are higher order
than the ones already dropped. Therefore, the achievement of an accuracy
better than the leading order requires the systematic
separation of the particle and antiparticle contributions, together with
the correct counting of the contributions from the ``flavor-changing''
term of Eq.~(\ref{fourquark}). We then postpone the explicit expression
of the reduced Lagrangian to the next section, where we discuss this 
more systematic nonrelativistic reduction.

\section{Nonrelativistic Effective Theory}
\label{NRred}

   The possibility of describing the high-$T$ QCD screening physics 
by a renormalizable local Lagrangian in 2+1 dimension, is valid only
up to one-loop level. However, since our specific goal is to study 
the screening physics, it still makes sense to give up the
renormalizability and derive an effective action for solving screening 
states in the high-$T$ limit with accuracy better than the leading 
order. In fact, the effective action necessarily 
contains higher dimensional operators. The coefficients of these 
higher dimensional operators will be derived by using the so-called 
matching technique, in close analogy with the nonrelativistic
reduction applied to the positronium~\cite{caswell} in QED and to 
heavy-quark systems~\cite{lepage} in QCD.

  In this section we derive a (2+1)-dimensional effective theory that 
describes screening states with accuracy up to $\tilde{g}^4(T)$,
apart from an overall additive zero-point energy which we determine
only up to $\tilde{g}^2(T)$.
As discussed at the end of the previous section, this expansion requires 
an explicit separation of the particle and antiparticle sectors,
because of the asymmetry of the mass-shell condition. Therefore,
this new theory, which improves results for screening mass 
splittings by one order in $\tilde{g}^2(T)$, becomes necessarily 
nonrelativistic-like for quarks.

The basic strategy is the following. We write down the most general
nonrelativistic Lagrangian, with terms up to some power in the
appropriate power counting scheme. Then the coefficients of these
terms are chosen so that they reproduce Green's functions of the
original theory expanded up to the same power in the coupling in
the relevant kinematic region.

\subsection{Notations}
Since we are interested in the regime of the reduced theory where 
quarks are close to their mass-shell, in the sense discussed in
Sec.~\ref{DR}, it is convenient to explicitly rotate from an Euclidean
notation (lower case letters) to a Minkowskian notation (upper case 
letters). We rotate the original first spatial direction to the 
time direction of the (2+1)-dimensional theory, while the other two 
spatial directions remain the spatial directions of the
(2+1)-dimensional theory: from now on, bold face letters indicate 
these two-dimensional vectors. We label the original time direction 
as the third axis, which now represents the chirally invariant mass
in the reduced theory. Specifically, the momentum of a
quark mode with tree-level mass $M=\pi T$ is rotated according to
$$(p_0=iM,p_1,p_2,p_3)\rightarrow((M+P_0)=-ip_1,P_1=p_2,P_2=p_3,M=p_0)
\equiv ((M+P_0),\bbox{P},M)\, .$$
The related rotation of the Dirac matrices is
$$(\gamma_0,\gamma_1,\gamma_2,\gamma_3)\rightarrow
(\Gamma_0=i\gamma_1,\Gamma_1=-\gamma_2,\Gamma_2=-\gamma_3,
\Gamma_3=-\gamma_0)\equiv(\Gamma_0,\bbox{\Gamma},\Gamma_3).$$ 
We use for the matrices $\Gamma$ the  explicit form of
Itzykson and Zuber~\cite{itzy}: 
$\Gamma_0=\sigma_3\otimes I_{2\times 2}$,
$\bbox{\Gamma}=i\sigma_2\otimes\bbox{\sigma}$, and
$\Gamma_3=i\sigma_2\otimes\sigma_3$. This choice of the $\Gamma$
matrices yields the following explicit form of the on-shell spinors, 
which are defined by $[E(\bbox{P})\Gamma_0
-\bbox{P}\cdot\bbox{\Gamma}-M\Gamma_3]U(\bbox{P})=0$
and $[E(\bbox{P})\Gamma_0
-\bbox{P}\cdot\bbox{\Gamma}+M\Gamma_3]V(\bbox{P})=0$,
\begin{equation}
U(\bbox{P})=
{1\over\sqrt{2}E(\bbox{P})}
\left(\begin{array}{c}
{\bbox{\sigma}\cdot\bbox{P}+\sigma_3 M} 
\\ E(\bbox{P}) \end{array}\right)\, ,
\quad\quad\quad
V(\bbox{P})=
{1\over\sqrt{2}E(\bbox{P})}
\left(\begin{array}{c}
\sigma_3E(\bbox{P}) \\ -\sigma_3\bbox{\sigma}\cdot\bbox{P}-M
\end{array}\right)\, ,
\end{equation}
where $E(\bbox{P})=\sqrt{M^2+\bbox{P}^2}$ is the on-shell energy
of a free quark in the reduced theory. These spinors specify
the basis in which the relativistic 4-component quark field is
decomposed as the nonrelativistic 2-component (spin up and down)
particle field and 2-component antiparticle field.

For convenience, we also rename the 
original gauge fields in the reduced theory as: $A_1={\cal A}_0$, 
$(A_2,A_3)=\bbox{\cal A}$ and $A_0=\varphi$. Since in the reduced 
theory $\varphi$ transforms like a matter field under the original 
static gauge transformation, it is often called the ``Higgs'' field.

\subsection{Dimensional analysis}

In general, the expansions in powers of $\tilde{g}^2(T)$ and in
derivatives (or low momenta) are two independent expansions. However,
when we consider bound states (screening states in our case) that
are dominated by the perturbative interaction, the two expansions
become intertwined. The reason is that the typical momentum is no longer
an independent variable, but rather it is determined by the interaction,
in contrast with scattering experiments where one controls the
momenta externally. If the interaction is weak, the typical momentum
is proportional to some power of $\tilde{g}^2(T)$.

It is possible to develop a systematic method for counting the
contribution of each term in powers of the coupling constant
to this combined expansion. The rationale of this 
method is described in detail in Ref.~\cite{lepage}, and it
is basically based on the analysis of the relevant
Schr\"{o}dinger equation with a potential derived from the 
tree-level approximation: in our case $V(\bbox{x})\sim g^2T\ln|\bbox{x}|$.
The resulting power-counting rules valid for studying screening states
(bound states of the reduced theory in 2+1 dimensions) are shown in
Table~\ref{table2}.

It is important to emphasize that the power counting is determined
by the leading behavior of the potential in terms of the coupling
constant. Therefore, it is still valid even when the potential 
acquires a nonperturbative linear confining term. In fact, in
the high-$T$ limit the perturbative tree-level potential overwhelms
the induced spatial string tension 
$\sigma_s(T)\propto \tilde{g}^4(T)T^2$~\cite{karsch}. 

Quarks in the reduced theory are very heavy making the time 
direction special relative to the spatial ones: this fact is
reflected in the form of the effective Lagrangian, which is
nonrelativistic, and also in the power counting rules of
Table~\ref{table2}

\subsection{Tree level}

At the tree level there are two kinds of corrections to the
tree-level NR Lagrangian: kinematics corrections and corrections
to the elastic scattering of a quark from external sources.

We do not need to consider inelastic scattering, because we
impose that the reduced theory contain only one power of the 
time derivative $\partial_t$: this requirement corresponds to the
precise choice of the field parameterization given in the $U$ and $V$
basis. There are no inelastic terms at the tree level in the
original theory, and the inelasticity only appears at the one-loop
level with this specific choice. In principle, one
can relate this particular choice of the field parameterization
to others that involve higher power of $\partial_t$ using the
invariance of the physics under field redefinition.

According to the power counting rules shown in Table~\ref{table2}, 
the only corrections up to order $\tilde{g}^4(T)$ relative to the
leading term are the following ones.\\
(a) Kinematics correction:
\begin{equation}
\overline{U}(\bbox{P})P\cdot\Gamma\, U(\bbox{P})
=M+P_0-E(\bbox{P})=P_0-{\bbox{P}^2\over 2M}
+{\bbox{P}^4\over 8M^3}\, .
\end{equation}
(b) Scattering from external ${\cal A}_0$:
\begin{eqnarray}
& &\overline{U}(\bbox{P})
\tilde{\cal A}_0(\bbox{P}-\bbox{P}')\Gamma_0\,
U(\bbox{P}')\nonumber \\
&=&\tilde{\cal A}_0(\bbox{P}-\bbox{P}')
\left[1-{1\over 4M^2}(\bbox{P}-\bbox{P}')^2
+{i\over 2M^2}\sigma_3\bbox{P}\times\bbox{P}'
+{i\over 2M}\bbox{\sigma}\times(\bbox{P}-\bbox{P}')\right]\, .
\end{eqnarray}
(c) Scattering from external $\bbox{\cal A}$:
\begin{equation}
\overline{U}(\bbox{P})
\tilde{\bbox{\cal A}}(\bbox{P}-\bbox{P}')\cdot
\bbox{\Gamma}\, U(\bbox{P}')
=\tilde{\bbox{\cal A}}_i(\bbox{P}-\bbox{P}')
\left[ {1\over 2M}(P_i+P_i')
-{i\over 2M}\epsilon_{ij}(P_j-P_j')\sigma_3\right]\, .
\end{equation}
(d) Scattering from external $\varphi$:
\begin{equation}
\overline{U}(\bbox{P})
\tilde{\varphi}(\bbox{P}-\bbox{P}')
\Gamma_3\, U(\bbox{P}')=
\tilde{\varphi}(\bbox{P}-\bbox{P}')
\left[1-{1\over 4M^2}(\bbox{P}^2+{\bbox{P}'}^2)
+{i\over 2M}\bbox{\sigma}\times(\bbox{P}-\bbox{P}')\right]\, .
\end{equation}

\subsection{One-loop level}
In Sec.~\ref{DRcalc} we have already calculated the heavy mode 
contributions at one loop. Since now we are using a NR 
formulation of the reduced theory in the quark sector, the light
quark contributions of the original theory are not exactly equal 
to the ones in the NR reduced theory, even if they have the same
infrared behavior. Therefore, we also need to calculate the 
one-loop contribution of the light quark modes both in the original
theory and in the NR reduced one: the difference between the two
results need to be added as a correction to the reduced Lagrangian
together with the heavy mode contributions.
In addition, the one-loop amplitudes needs to be sandwiched with
the appropriate spinors to yield the correct correction terms.
The calculation is straightforward and we only list the resulting
terms up to $\tilde{g}^4(T)$.

The nonrelativistic one-loop corrections, which need to be compared
to the analogous corrections of the light quarks in the original
theory, are calculated using the following tree-level Lagrangian
\begin{equation}
{\cal L}^{NR}_{tree}=
\psi^{\dag}\left[iD_t+{\bbox{D}^2\over 2M}
-g_3\varphi\right]\psi \, ,
\end{equation}
where $D_t\equiv\partial_t+ig_3{\cal A}_0$,
$\bbox{D}\equiv\bbox{\partial}-ig_3{\cal A}$ and $\psi$
now is a two-component (representing spin up and down)
nonrelativistic quark field.
One can easily work out the Feynman rules associated with
this Lagrangian and calculate  the relevant one-loop graphs.
Again, we only list final results, which have been computed using
dimensional regularization.

In the following the one-loop corrections to the original theory
both from heavy and light quarks are grouped together (the ones coming
from sandwiching with appropriate spinors, the terms calculated in
Sec.~\ref{DRcalc} and the ``new'' ones coming from the lowest Matsubara 
frequency). The one-loop contributions from the NR reduced theory are
instead given separately. The explicit values of the $X$'s, $Y$'s and
$Z$'s are given in Table~\ref{table2}.\\
(e) Self-energy corrections:
\begin{eqnarray}
\label{secr}
& &\overline{U}(\bbox{P})i\Sigma\, U(\bbox{P})=
C_f \,{\cal G}^2(T)\Biggl\{-4(M-P_0) \nonumber \\
& &\quad\quad\quad\quad+\biggl[2M-(X+X_{11}-iX_{10}+2c)P_0
+(X-X_{00}-iX_{10}+2c){\bbox{P}^2\over 2M}\biggr]\Biggr\}\, . \\
& &i\Sigma^{NR}=C_f \,{\cal G}^2(T) \,
\Biggl\{-2M\biggl[{1\over\epsilon}-\gamma_E
+\ln{\pi\mu^2\over M^2}\biggr]
-P_0+{3\over 4}{\bbox{P}^2\over M}\Biggr\}\, .
\label{secnr}
\end{eqnarray}
The momentum independent divergence in $i\Sigma^{NR}$ is related to 
the additive mass renormalization in the nonrelativistic Lagrangian. 
This additive mass term is often explicitly ignored at the price of
introducing a zero-point energy ambiguity in the nonrelativistic theory,
i.e. we can only calculate mass differences. In principle this ambiguity
can be resolved at least perturbatively in each specific regularization
scheme.\\
(f) Abelian-like vertex corrections:
\begin{eqnarray}
& &\overline{U}(\bbox{0})\Delta\Gamma_0\, U(\bbox{0})
=i\tilde{g}(T){\lambda^a\over 2}
\Bigl(C_f-{C_{ad}\over 2}\Bigr)\, {\cal G}^2(T)
\Biggl\{-4+\Bigl[Y+Y_{11}-iY_{10}+2c\Bigr]\Biggr\}\, . \\
& &\overline{U}(\bbox{0})\Delta\Gamma_3\, U(\bbox{0})
=i\tilde{g}(T){\lambda^a\over 2}
\Bigl(C_f-{C_{ad}\over 2}\Bigr)\, {\cal G}^2(T)
\Biggl\{-4+\Bigl[Y+Y_{00}+iY_{01}+2c\Bigr]\Biggr\}\, . \\
& &\Delta\Gamma_0^{NR}=\Delta\Gamma_3^{NR}
=-i\tilde{g}(T){\lambda^a\over 2}
\Bigl(C_f-{C_{ad}\over 2}\Bigr)\,2{\cal G}^2(T)\, .
\end{eqnarray}
(g) Nonabelian-like vertex corrections:
\begin{eqnarray}
& &\overline{U}(\bbox{0})\Delta\Gamma_0\, U(\bbox{0})
=i\tilde{g}(T){\lambda^a\over 2}\,C_{ad}\,{\cal G}^2(T)\,
\Biggl\{-2+\Bigl[Z+Z_{11}-iZ_{10}+3c\Bigr]\Biggr\}\, . \\
& &\overline{U}(\bbox{0})\Delta\Gamma_3\, U(\bbox{0})
=i\tilde{g}(T){\lambda^a\over 2}\,C_{ad}\,{\cal G}^2(T)\,
\Bigl[Z+Z_{00}+iZ_{01}+3c\Bigr]\, . \\
& &\Delta\Gamma_0^{NR}=\Delta\Gamma_3^{NR}=0 \, .
\end{eqnarray}
Since the leading terms are already order $\tilde{g}^3(T)$, the
momentum dependent terms are higher orders and can be dropped in
vertex corrections.

As expected, the one-loop corrections in the original theory and 
the NR reduced theory do not match exactly. The differences can
be compensated up to order $\tilde{g}^4(T)$ by adding the following
terms to the (2+1)-dimensional NR Lagrangian 
\begin{equation}
\Delta{\cal L}_{1-loop}={\cal G}^2(T)\,\biggl\{
c_1\, \psi^{\dag}i\partial_t\psi+
c_2\, \psi^{\dag}{\bbox{\partial}^2\over 2M}\psi-
c_3\, \psi^{\dag}g_3{\cal A}_0\psi-
c_4\, \psi^{\dag}g_3\varphi\psi \biggr\}\, .
\label{delta_L}
\end{equation}
The coefficients $c_i$'s, which are determined by subtracting the
one-loop corrections obtained in the NR reduced theory from those
computed in the original theory, are given in Table~\ref{table3}.

It is important to realize that we have taken out a factor of 2
from the one-loop results computed in the original theory before 
we subtract the nonrelativistic results from them. The necessity 
of dividing out this factor of 2 comes from the fact that the
Lagrangian in Eq.~(\ref{qcd_3d}) has two degenerate ``flavors''
(with $M=\pm \pi T$),
whereas the NR one-loop results are for a single ``flavor''.
If the physics requires a normal flavor structure, the relevant
flavor content must be added explicitly to the NR Lagrangian.

\subsection{Reduced Lagrangian: fermionic part}
We can now write down the final result for the fermionic part of 
the DR Lagrangian by collecting the corrections from the last
two subsections and properly gauging each derivative.

The final form of this Lagrangian is more suggestive if written in
terms of the appropriate color electric and magnetic fields ($i=1,2$)
\begin{equation}
  \bbox{\cal E}_i \equiv {\cal F}_{0i}\, ,
  \quad\quad
  {\cal B}_3 \equiv {1\over 2}\epsilon_{ij3}{\cal F}_{ij}\, ,
\end{equation}
where 
${\cal F}_{ij}=\partial_i{\cal A}_j-\partial_j{\cal A}_i
-ig_3[{\cal A}_i,{\cal A}_j]$
is the gauge field strength tensor in 2+1
dimensions. Note that there is only one component of the color 
magnetic field in 2+1 dimensions. The final Lagrangian is given by
\begin{equation}
{\cal L}_{F} = {\cal L}^{(0)}+{\cal L}^{(1s)}+{\cal L}^{(2)}
+{\cal L}^{(2s)}\, ,
\label{L_reduced}
\end{equation}
with
\begin{mathletters}
\label{L_redpart}
\begin{eqnarray}
{\cal L}^{(0)} &=&
\Psi^{\dag}\biggl( i D_t + \frac{\bbox{D}^2}{2M}\biggr)\Psi 
   -g_\varphi \Psi^{\dag} \varphi \Psi\, , \\
{\cal L}^{(1s)} &=&
   - \frac{g_3}{2M} \Psi^{\dag} \bbox{\sigma}
     \times \bbox{\cal E}  \Psi 
  + \frac{g_\varphi}{2M} \Psi^{\dag} \bbox{\sigma}
     \times \Bigl(\bbox{D}\varphi\Bigr) \Psi \\
{\cal L}^{(2)} &=&
\frac{g_3}{4M^2} \Psi^{\dag}\Bigl( \bbox{D}\cdot \bbox{\cal E}
        - \bbox{\cal E} \cdot  \bbox{D} \Bigr)\Psi  
 - \frac{g_\varphi}{4M^2} \Psi^{\dag} \{\bbox{D}^2,\varphi \}\Psi 
 + \Psi^{\dag} \frac{\bbox{D}^4}{8M^3} \Psi \\
{\cal L}^{(2s)} &=&
\frac{g_3}{2M} \Psi^{\dag} {\sigma}_3{\cal B}_3 \Psi
 +  \frac{i g_3}{4M^2} \Psi^{\dag} \sigma_3
    \Bigl( \bbox{D} \times \bbox{\cal E} -
           \bbox{\cal E} \times \bbox{D} \Bigr) \Psi \, ,
\end{eqnarray}
\end{mathletters}
where the sum over the color indices is implicit.
In the above equation we have absorbed the one-loop corrections,
due to $\Delta{\cal L}_{\text{1-loop}}$ in Eq.~(\ref{delta_L}),
into finite renormalization of the proper physical quantities:
\begin{mathletters}
\begin{eqnarray}
\Psi &=& \Bigl[1+{c_1\over 2} {\cal G}^2(T)\Bigr]\psi\, ,\\
M &=& \pi T\,\Bigl[1+(c_1-c_2) {\cal G}^2(T)\Bigr]\, , \\
g_3 &=&\tilde{g}(T)\sqrt{T}\Bigl[1-(c_1-c_3) {\cal G}^2(T)\Bigr]\, , \\
g_\varphi &=&\tilde{g}(T)\sqrt{T}\Bigl[1-(c_1-c_4) {\cal G}^2(T)\Bigr]\, .
\end{eqnarray}
\label{param}
\end{mathletters}

At this point it is appropriate to make some general remarks
concerning the fermionic effective Lagrangian.\\
(1) The Lagrangian in Eqs.~(\ref{L_reduced}) and 
(\ref{L_redpart}) has contribution only from
a ``single-flavor'' particle state. The reduced Lagrangian for
the antiparticle state has the same form except that couplings 
change sign. The four-fermion term in Eq.~(\ref{fourquark}) does
not give contribution to order $\tilde{g}^4(T)$ because of the
separation of the quark and antiquark sectors.\\
(2) According to the power counting rule in Table~\ref{table2}, 
${\cal L}^{(0)}$ begins to contribute to the binding energy at 
order $\tilde{g}^2(T)T$, ${\cal L}^{(1s)}$ at order $\tilde{g}^3(T)T$,
and ${\cal L}^{(2)}$ and ${\cal L}^{(2s)}$ at order $\tilde{g}^4(T)T$.
While ${\cal L}^{(0)}$ and ${\cal L}^{(2)}$ are spin independent,
${\cal L}^{(1s)}$ and ${\cal L}^{(2s)}$ are spin-dependent. \\
(3) Even though there exist NR four-fermion terms that are order
$\tilde{g}^4(T)$ according to a naive application of the power-counting
rules of Table~\ref{table2}, e.g. $g^2(\Psi^{\dag}\Psi\Phi^{\dag}\Phi)/M$ 
($\Phi$ is the antiquark field), these terms can only contribute 
to the binding energy through higher-loop graphs and, therefore, their
contributions are in fact of order higher than $\tilde{g}^4(T)$.  \\
(4) Notice that ${\cal L}^{(1s)}$ is absent in 3+1 dimensions
since it is not a scalar under spatial rotation, while it is a scalar
under a two-dimensional rotation around the 3rd direction.\\
(5) Since the self-energy in the NR theory is specified up to an
additive constant, the Lagrangian in Eq.~(\ref{L_reduced}) cannot
give the zero-point energy. However, it is still possible to 
determine perturbatively the zero-point energy shift between the NR
Lagrangian and the original one. For instance, the formulae
given in Eqs.~(\ref{secr}) and (\ref{secnr}) give the zero-point 
energy shift up to order $\tilde{g}^2(T)$. If we want to reach the same
accuracy we have obtained for the energy differences (splittings),
i.e. $\tilde{g}^4(T)$, also for the zero-point energy shift we need to
perform a two-loop calculation of the  quark self-energy.\\
(6) The fact that ${\cal G}^2(T)$ is small, as we shall see shortly,
means that corrections from heavy
modes and, therefore, the coefficients of higher dimensional
operators are small and can be calculated perturbatively. It does
not mean that the physics governed by the DR Lagrangian, whose
coupling constant $\tilde{g}^2 (T) T$ is very large at high $T$,
is perturbative. In fact, the infrared behavior of the DR Lagrangian,
by construction, remains the same as the original theory.\\
(7) The coefficients in Eq.~(\ref{L_reduced}) has been derived in the
$\overline{\text{MS}}$ scheme. If one wants to solve the reduced
theory in a different renormalization scheme, e.g. on the lattice,
one needs, in principle,  to compute again these coefficients
in that specific scheme. In practice, this may not be necessary,
since ${\cal G}^2(T)$ has turned out to be numerically very
small and, hence, the tree-level coefficients, which are scheme 
independent, dominate. One could have problems only in those scheme
that have corrections anomalously large.\\
(8) The full gauge invariance in the original theory is now reduced
to the ``static'' gauge invariance, which is explicitly kept in the
above Lagrangian.\\
(9) In analogy to universality in the pure bosonic case, a similar 
universality also holds in the quark sector. After the one-particle 
irreducible graphs have been matched in the relevant kinematic region 
and the reduced Lagrangian obtained, any other Green's function can be 
computed from this same Lagrangian to within the same accuracy of
the matching. For example, we expect that the reduced theory is capable 
of describing the screening states, e.g. those states that describe
the spatial correlation of mesonic currents; these states can be interpreted
as bound states in the reduced world and hence they are beyond those 1PI 
graphs we have used to derive the reduced theory. We have explicitly 
verified this fact in the Gross-Neveu model~[14].
In particular, the screening mass splittings can be solved from the
reduced theory to an accuracy of $\tilde{g}^4(T)$, though only
nonperturbatively.

\subsection{Reduced Lagrangian: bosonic part}
For the purpose of describing mesonic and baryonic screening
states to order $\tilde{g}^4(T)$, we only need the bosonic part of the
Lagrangian ${\cal L}_B$ to order $\tilde{g}^2(T)$. In fact, these screening
states reduce to two (three) valence quarks in infinite weak coupling
(infinite $T$) limit and, therefore, the gluonic contribution to the 
binding coming from ${\cal L}_B$ has to involve one more loop and hence
an additional $\tilde{g}^2(T)$ factor.
For this reason, it is sufficient for our purpose to use the bosonic 
part of the reduced Lagrangian up to one loop, which has 
already been derived in Refs.~\cite{landsman,kajantie}
\begin{equation}
{\cal L}_B=-{1\over 4}{\cal F}_{\mu\nu}^a{\cal F}_{\mu\nu}^a
-{1\over 2}(D_\mu\varphi)^a(D_\mu\varphi)^a
-{1\over 2}m_D^2\, \varphi^a\varphi^a+\ldots\, ,
\label{L_B}
\end{equation}
where $m_D^2=\tilde{g}^2(T)(N/3+N_f/6)T^2$ is the one-loop Debye
screening mass, $\mu,\nu=0,1,2$ and the dots represent terms that
contribute to the binding energy at orders higher than $\tilde{g}^4(T)$,
e.g. the interaction terms between $\varphi$ and ${\cal A}_\mu$.

\subsection{The complete reduced Lagrangian}
   In summary, the complete reduced theory that should describe
mesonic and baryonic screening physics with accuracy up to 
$\tilde{g}^4(T)$ for mass splittings and up to $\tilde{g}^2(T)$ 
for the overall mass is given by the sum of quark DR Lagrangian
${\cal L}_F$ in Eq.~(\ref{L_reduced}), the corresponding antiquark 
DR Lagrangian, which is still given by Eq.~(\ref{L_reduced}) with opposite
sign for the couplings, and the gluonic Lagrangian ${\cal L}_B$ 
in Eq.~(\ref{L_B}). The correspondence between the parameters in QCD
and in the reduced theory is given explicitly in Eq.~(\ref{param}),
and in Eqs.~(\ref{coupling}--\ref{newcoupling}).
In the kinematic region that is relevant for these screening states,
i.e. those states that dominate the large-distance correlation between 
currents, the lightest quark mode is close to its mass-shell, and the
self-consistency of the reduced theory is guaranteed by the power counting 
rule given in Table~\ref{table2}, which have been justified in 
Refs.~\cite{caswell,lepage}, exactly in parallel with the heavy quarkonium 
systems. The essence of the argument is that, if the interaction is weak, the 
off-shellness is small.

  A final word of caution about the usefulness of this reduced theory.
Thermodynamic quantities, such as the free energy, specific heat and 
so on, are dominated by the gluonic zero modes. As it is well-known,
quarks are heavy relative to gluonic zero modes and hence their
contribution to thermodynamic quantities is
suppressed at high $T$. This is perhaps the
reason why quark degrees of freedom have never been considered relevant
at high $T$, until one has been explicitly interested in the mesonic and 
baryonic screening states. The effective theory we derived is not intended 
for bulk thermodynamic observables, but only for those screening 
states that do not mix with purely gluonic states.

\subsection{At what temperature do we expect DR?}

At last we have obtained a reduced theory that includes the quark 
sector and that should be valid in the high-$T$ limit. However, 
we still face the practical question of estimating the temperature
above which this DR theory is going to be a good approximation to QCD.

We know that DR is manifest only in a limited class of subtraction
schemes~\cite{landsman}, but even within this class there is some
freedom left: one aspect of this freedom is the choice of the
coupling constant in the reduced theory $g_3^2\equiv\tilde{g}^2(T)T$.
While it is unambiguous how $\tilde{g}^2(T)$ runs with $T$, the
numerical value of $\tilde{g}^2(T)$ at a specific temperature
depends on how we match the reduced and the original theory.
For instance, we have argued that a physically relevant way of
choosing the coupling in the pure gluonic sector is to match the 
effective actions in the background field scheme~\cite{lambda}.
This specific choice yields that DR in the pure glue sector sets in
around $2T_c$, i.e. the coupling constant is sufficiently small at 
that temperature. 

This freedom in the choice of the relevant
coupling constant can be rephrased in our formalism with the
freedom of choosing the scale $\mu$ in Eq.~(\ref{scale_mu}) 
where $c$ has different values in different schemes.
Since our present interest is to solve screening states of quarks,
it might be better to choose the subtraction parameter $c$ with the
criterion that the one-loop couplings $g_3$ and $g_\varphi$ be
as close as possible to their tree-level values. For example, 
if we demand $c_1-c_3=0$, we get $c=-1/3$ (if $N=3$). This choice
would make the subtraction scale $\mu$ defined in 
Eq.~(\ref{scale_mu}) larger than the one obtained from the criterion 
of optimal DR for gluons, which yields instead $c=1/22$ 
($N=3$ and $N_f=0$). Alternatively, if we demand $c_1-c_4=0$, 
we find a even larger subtraction scale $\mu$ ($c=-1.455$).

In either case, the fact that the subtraction scale is larger 
makes the effective coupling constant at a given temperature smaller 
in the quark sector than in the pure gluonic sector. This result is 
certainly in qualitative agreement with the empirical fact that the 
DR in the quark sector sets in at temperatures almost right above the 
critical point~\cite{bs-lattice}, whereas DR in the pure
gluon sector sets in at about $T\approx 2T_c$~\cite{reisz,karsch}.

\section{Summary and Conclusions}
\label{conclu}
   We have shown with an explicit one-loop calculation that QCD at 
high temperature undergoes dimensional reduction also in the
quark sector. More specifically, we have
shown that static one-particle irreducible graphs contributing to the 
original four-dimensional correlation functions can be reproduced by a 
(2+1)-dimensional renormalizable Lagrangian to order $\tilde{g}^2(T)$ in
the kinematic region where the lowest Matsubara quark modes are close
to their ``mass-shell''. Physical reasons why this kinematic region
is relevant to screening phenomena have been also discussed.
The reduced theory only contains the zero modes of the gauge field and 
the lowest quark modes as the explicit degrees of freedom and has the
form given by Eq.~(\ref{qcd_3d}) plus the four-fermion term of 
Eq.~(\ref{fourquark}).

   Aiming at a better description of the mesonic and baryonic screening 
states, we have further improved the reduced theory to order
$\tilde{g}^4(T)$ via a nonrelativistic reduction, which results in the
reduced effective Lagrangian of Eqs.~(\ref{L_reduced}) and 
(\ref{L_B}). In fact, while the relativistic version of the reduced 
theory mixes different orders in the coupling when used in the kinematic 
region relevant to screening physics, the nonrelativistic reduction
explicitly separates the contribution of particles and antiparticles
and allows a correct counting of the expansion parameter.

  Furthermore, we have also argued that the reduced theory in the quark
sector, i.e. for screening mesonic and baryonic correlators that do not
mix with entirely gluonic states, should become
accurate at temperatures slightly above the chiral restoration transition
temperature, due to the smallness of the appropriate running coupling 
${\cal G}^2(T)$. In particular, we find that the temperature above which 
the reduced theory becomes reliable in the quark sector should be even 
lower than the corresponding temperature in the pure gluonic 
sector~\cite{lambda}. Our result has the potential for explaining 
present lattice data~\cite{mass-s,bs-s} and provides a formal basis to
the recent phenomenological modeling~\cite{bs-lattice} of the same data.

  We would like to stress that, although the reduced Lagrangian has
been derived in a perturbative context, the reduced theory embodies all
the infrared physics of the original theory, i.e. QCD at high temperature.
Therefore, the solution of the reduced theory should reproduce the
full long-wavelength screening physics of QCD in the high-$T$ limit,
which is nonperturbative: nonperturbative approaches such
as lattice simulations are required to find this solution. Luckily
since the large scale ($T$) has been explicitly factored out, it is now
straightforward to put the nonrelativistic version of the reduced theory 
on a lattice following, for example, the method of Ref.~\cite{lepage}.

\acknowledgments
  One of us (ML) gratefully thanks the Department of Physics 
at the University of Washington for their hospitality.
  This work was supported in part by funds provided by the U.S.
Department of Energy (DOE) under contract number DE-FG06-88ER40427
and cooperative agreement DE-FC02-94ER40818.

\begin{table}
\caption[]{One-loop coefficients. The entries that are not listed, i.e.
those coefficients with at least one index equal to 2 or 3, are zero.}
\begin{tabular}{ccr}
coefficient & analytic expression & numerical value \\
\tableline
$X$  & $2\gamma_E-24\zeta'(-1)-(14/3)\ln2-1$ & $0.8898$ \\
$X_{00}$& $X+1$ & $1.8898$ \\
$X_{01}$& $i[2\gamma_E-72\zeta'(-1)-10\ln2-6]$ & $i\,0.1332$ \\
$X_{10}$& $i[X-1]$ & $-i\,0.1102$ \\
$X_{11}$& $i\, X_{01}$ & $-0.1332$ \\
\tableline
$Y$  & $X$ & $0.8898$ \\
$Y_{00}$& $2\gamma_E+24\zeta'(-1)+(2/3)\ln2+4$ & $1.6465$  \\
$Y_{01}$& $X_{10}$ & $-i\,0.1102$ \\
$Y_{10}$& $X_{10}$ & $-i\,0.1102$ \\
$Y_{11}$& $X_{11}$ & $-0.1332$ \\
\tableline
$Z$ & $-2\gamma_E+12\zeta'(-1)+(10/3)\ln2+1/2$ & $-0.3290$ \\
$Z_{00}$& $\gamma_E-12\zeta'(-1)-(7/3)\ln2$ & $0.9449$ \\
$Z_{01}$& $i[12\zeta'(-1)+(4/3)\ln2+1]$ & $-i\,0.0609$ \\
$Z_{10}$& $Z_{01}$ & $-i\,0.0609$ \\
$Z_{11}$& $\gamma_E-36\zeta'(-1)-5\ln2-3$ & $0.0666$\\
\tableline
$W$ & $\sum_{n=2}^{\infty} 
{1\over n} \ln \left[ 
   {(4n^2-1)^2 \over 16 n^2 (n^2-1)}
             \right]$  & $0.1205$ \\
$W_{00}$& $-2\gamma_E+24\zeta'(-1)+(17/3)\ln2+2\ln3 + W$ & $1.1210$ \\
$W_{01}$& $-3\ln2+2\ln3+W$ & $0.2383$\\
$W_{11}$& $-24\zeta'(-1)-(35/3)\ln2+2\ln3+2+ W$ & $0.2011$
\end{tabular}
\label{table1}
\end{table}

\begin{table}
\caption[]{Power counting rules in the kinematic region appropriate
for studying screening states at high $T$.}
\begin{tabular}{ccccccccccc}
$\epsilon_B$ & $K$ & $\bbox{P}$ & $\psi$ & $i\partial_t$ &
$i\bbox{\partial}$ &$ g_3{\cal A}_0$ & $g_3\bbox{\cal A}$ &
$g_3\varphi$ & $g_3\bbox{\cal E}$ & $g_3\bbox{\cal B}$ \\ \hline
$g^2T$ & $g^2T$ & $gT$ & $gT$ & $g^2T$ & $gT$ & $g^2T$ & $g^3T$ &
$g^2T$ & $g^3T^2$ & $g^4T^2$ 
\end{tabular}
\label{table2}
\end{table}

\begin{table}
\caption[]{Coefficients of the one-loop corrections.}
\begin{tabular}{ccr}
coefficient & analytic expression & numerical value \\ \hline
$c_1$ & $C_f[c-3+(X+X_{11}-iX_{10})/2]$ & $C_f[c-2.6768]$ \\
$c_2$ & $C_f[c-(3-X+X_{00}+iX_{10})/2]$ & $C_f[c-2.0551]$ \\
$c_3$ & $C_f[c+(Y+Y_{11}-iY_{10})/2]+C_{ad}[c-1]$ &
$C_f[c+0.3232]+C_{ad}[c-1]$ \\
$c_4$ & $C_f[c+(Y+Y_{00}+iY_{01})/2]$ & $C_f[c+1.3232]$ \\
 & $+C_{ad}[c-(Y+Y_{00}+iY_{01})/4+(Z+Z_{00}+iZ_{01})/2]$ &
$+C_{ad}[c-0.3232]$
\end{tabular}
\label{table3}
\end{table}

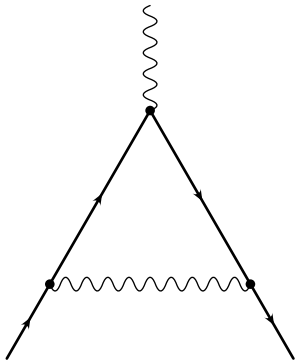
\end{document}